\documentclass[acmlarge]{acmart}

\pdfoutput=1
\usepackage{float}
\usepackage[ruled]{algorithm2e}
\usepackage[section]{placeins}
\usepackage{tabu}

\usepackage{caption}

\usepackage{bigfoot}

\usepackage{subcaption}

\usepackage{array}
\usepackage{afterpage}

\definecolor{orangered2}{HTML}{FA4616}
\definecolor{mediumblue}{HTML}{0000CD}
\definecolor{pank}{HTML}{BF3EFF}
\definecolor{darkgreen}{HTML}{006400}

\usepackage{multirow}
\usepackage{datetime}
\usepackage{amsmath}
\usepackage{diagbox}
\usepackage{listings}
\usepackage{xcolor}
\usepackage[utf8]{inputenc}
\usepackage[multiple]{footmisc}

\usepackage{etoolbox}
\makeatletter
\patchcmd{\maketitle}{\@copyrightspace}{}{}{}

\newcommand{\ooorange}[1]{{\textbf{\color{orangered2}#1}}}

\newcommand{\greeeeen}[1]{{\textbf{\color{pank}#1}}}
\newcommand{\brown}[1]{{\textbf{\color{darkgreen}#1}}}

\newcommand{\URIR}{\mbox{URI-R} }
\newcommand{\URIM}{\mbox{URI-M} }

\newcommand{\URIRs}{\mbox{URI-Rs} }
\newcommand{\URIMs}{\mbox{URI-Ms} }

\newcommand{\URIMns}{\mbox{URI-M}}
\newcommand{\URIRns}{\mbox{URI-R}}

\newcommand{\URIRsns}{\mbox{URI-Rs}}
\newcommand{\URIMsns}{\mbox{URI-Ms}}

\usepackage{lmodern}
\usepackage[T1]{fontenc}
\usepackage[utf8]{inputenc}

\usepackage{wrapfig}
\usepackage{rotating}
\usepackage{balance}
\usepackage{morefloats}
\usepackage{fancyvrb,cprotect}

\usepackage{hyperref} 

\def\hreffootnote{%
    \begingroup
    \def\Url@HyperHook ##1\endgroup{%
        \let\Url@HyperHook\relax
        \href@footnote
    }%
    \url
}

\let\realhref\href
\def\href@footnote#1\endgroup#2{%
   \footnote{\expandafter\realhref\expandafter{\Url@String}{#2}}%
   \endgroup\endgroup#1\endgroup
}

\acmArticle{-1}

\setcopyright{none}
\renewcommand\footnotetextcopyrightpermission[1]{}
\begin{document}

\title{Impact of URI Canonicalization on Memento Count}

\author{Mat Kelly}
\orcid{0000-0002-0236-7389}
\email{mkelly@cs.odu.edu}
\affiliation{%
  \institution{Old Dominion University}
  \department{Department of Computer Science}
  \city{Norfolk}
  \state{Virginia}
  \country{USA}
}
\author{Lulwah M. Alkwai}
\email{lalkwai@cs.odu.edu}
\affiliation{%
  \institution{Old Dominion University}
  \department{Department of Computer Science}
  \city{Norfolk}
  \state{Virginia}
  \country{USA}
}
\author{Michael L. Nelson}
\orcid{0000-0003-3749-8116}
\email{mln@cs.odu.edu}
\affiliation{%
  \institution{Old Dominion University}
  \department{Department of Computer Science}
  \city{Norfolk}
  \state{Virginia}
  \country{USA}
}
\author{Michele C. Weigle}
\orcid{0000-0002-2787-7166}
\email{mweigle@cs.odu.edu}
\affiliation{%
  \institution{Old Dominion University}
  \department{Department of Computer Science}
  \city{Norfolk}
  \state{Virginia}
  \country{USA}
}

\author{Herbert Van de Sompel}
\affiliation{
  \institution{Los Alamos National Laboratory}
  \city{Los Alamos}
  \state{New Mexico}
  \country{USA}
}
\email{herbertv@lanl.gov}

\begin{abstract}
Quantifying the captures of a URI over time is useful for researchers to identify the extent to which a Web page has been archived. Memento TimeMaps provide a format to list mementos (URI-Ms) for captures along with brief metadata, like Memento-Datetime, for each \URIMns. However, when some \URIMs are dereferenced, they simply provide a redirect to a different URI-M (instead of a unique representation at the datetime), often also present in the TimeMap. This infers that confidently obtaining an accurate count quantifying the number of non-forwarding captures for a \URIR is not possible using a TimeMap alone and that the magnitude of a TimeMap is not equivalent to the number of representations it identifies. In this work we discuss this particular phenomena in depth. We also perform a breakdown of the dynamics of counting mementos for a particular \URIR (google.com) and quantify the prevalence of the various canonicalization patterns that exacerbate attempts at counting using only a TimeMap. For google.com we found that 84.9\% of the URI-Ms result in an HTTP redirect when dereferenced. We expand on and apply this metric to TimeMaps for seven other URI-Rs of large Web sites and thirteen academic institutions. Using a ratio metric $DI$ for the number of \URIMs without redirects to those requiring a redirect when dereferenced, five of the eight large web sites' and two of the thirteen academic institutions' TimeMaps had a ratio of ratio less than one, indicating that more than half of the \URIMs in these TimeMaps result in redirects when dereferenced.\vspace{-1.5em}
\end{abstract}

\maketitle

\section{Introduction}
\label{sec:intro}
Memento TimeMaps serve as an index for the mementos for an original resource (\URIRns) contained in an archive \cite{rfc7089}. Web archives return TimeMaps with a list of identifiers (URI-Ms) for the HTTP transactions observed at archival time. TimeMaps have generally been used as a count of the number of representations of a \URIR present in an archive. But, TimeMaps may include \URIMs for archived representations (HTTP 2XX), archived redirections (HTTP 3XX), and archived errors (HTTP 4XX or 5XX) \cite{rfc7231}. Only the URI-Ms that result in an HTTP 2XX when dereferenced match the general notion of a ``capture'' of the contents of a webpage. But, the status that results when a \URIM is dereferenced (to the extent of returning an archived entity body) is not present in a TimeMap. Further, TimeMaps do not explicitly return a ``count'' value to indicate the number of mementos listed in the TimeMap that produce a non-redirecting (non-3XX) HTTP status code when dereferenced.  This can cause problems when using the number of \URIMs in a TimeMap as a proxy for the number of captures of a Web page.

Various tools \cite{memgator,  jordan-jcdl2015, kelly-dl2014-mink} and access points into the Web archives return a different count of the number of captures for a \URIR depending on the heuristic implemented and the source of the archival listings. For example, the Internet Archive's web interface when queried with the URI-R example.com states, ``Saved 11,771 times between January 20, 2002 and May 20, 2016''. The file returned from Internet Archive's CDX endpoint returns 69,162 entries. The TimeMap from Internet Archive for example.com contains 40,641 URI-Ms with a rel value of ``memento''. The heuristic of determining how many captures are represented by URI-Ms in a TimeMap cannot be completed without dereferencing. 

Researchers can use the inline metadata about the \URIMns, without the need to dereference the \URIM in a TimeMap, including its temporal ordering, datetime (through the datetime HTTP Link attribute \cite{rfc7089}), etc. to infer characteristics about a dereferenced memento. However, dereferencing some URI-Ms in a TimeMap produces an HTTP redirect \cite{rfc7231} that instructs the client to access a \URIM with a different datetime, to obtain the requested content. For example, a TimeMap for \url{http://vimeo.com} from Internet Archive contained 199,262 URI-Ms with an associated ``rel'' value of ``memento''. However, when a user accesses over 57\% of these \URIMsns, an HTTP Redirect is returned pointing to another memento whose \URIM is in the TimeMap that returns a HTTP Status OK. A different extreme of memento count results when a user accesses the TimeMap from \url{http://odu.edu}, whose percentage of redirects is around 9.7\% of the URI-Ms listed.

Redirection in a Web archive can be attributed to a variety of canonicalization rules including a scheme change (e.g., http to https), an obsolete subdomain (e.g., www2 to www), a slash added to a URI (\hyperref[]{http://foo.com/\textasciitilde joe} to \hyperref[]{http://foo.com/\textasciitilde joe/}), among others. Preserving and replaying these redirects allows an archive to accurately reproduce the HTTP transactions that would have occurred when the URI being accessed resided on the live Web.

When a \URIM in a TimeMap is dereferenced, it may redirect to another \URIM listed in the TimeMap. Because of this, the heuristic of counting URI-Ms with relation values of ``memento'' is an inaccurate means of determining the number of unique representations inferred from a TimeMap. We further emphasize the distinction per the Memento specification that the identifiers for mementos (URI-Ms) in a TimeMap are identifiers for archived HTTP transactions (e.g., transmission of HTTP 2XX, 3XX, 4XX, etc.) rather than identifiers for representations.

Based on the number of URI-Ms in a TimeMap not necessarily resolving to unique mementos when archival redirects are followed, we examined the mementos from contemporarily large TimeMaps to evaluate the patterns and schemes used in Memento canonicalization. Through this, we identify the difference between the number of mementos available as reported by the TimeMap through naive ``rel'' counting heuristics to the temporally unique mementos identified once these mementos are dereferenced.

\section{Background}
\label{sec:bg}

This section includes background information and an overview of the state-of-the-art of archival technologies relevant to this work including Memento aggregation, URI canonicalization, archival indexing formats, and \URIR opacity.

\subsection{Memento Aggregation}

\begin{figure}
\begin{lstlisting}
<http://example.com>; rel="original",
<http://web.archive.org/web/20020120142510/http://example.com/>; rel="first ^memento^"; datetime="Sun, 20 Jan 2002 14:25:10 GMT",
<http://web.archive.org/web/20020804094019/http://www.example.com/>; rel="^memento^"; datetime="Sun, 04 Aug 2002 09:40:19 GMT",
<http://web.archive.org/web/20160728014649/http://www.example.com/>; rel="^memento^"; datetime="Thu, 28 Jul 2016 01:46:49 GMT",
<http://web.archive.org/web/20160728114745/http://www.example.com>; rel="^memento^"; datetime="Thu, 28 Jul 2016 11:47:45 GMT",
<http://web.archive.org/web/20160728123024/http://example.com/>; rel="last ^memento^"; datetime="Thu, 28 Jul 2016 12:30:24 GMT",
<http://localhost:1208/timemap/link/http://example.com>; anchor="http://example.com"; rel="timemap"; type="application/link-format",
<http://localhost:1208/timegate/http://example.com>; anchor="http://example.com"; rel="timegate"
\end{lstlisting}
\caption{A partial Link formatted TimeMap from a local instance of MemGator. Highlighted rel values constitute inclusion in the sum described in Equations~\ref{eq:naive} and \ref{eq:naive2}.}
\label{fig:tm_link}
\end{figure}

The Memento Framework \cite{rfc7089} allows navigation of Web archives in the dimension of time using content from Web archives and resource versioning systems. A Memento TimeMap (Figure~\ref{fig:tm_link}) is a structured list of identifiers (URI-Ms) for archived captures (mementos) returned from an archive when queried with a \URIR as the parameter. A TimeMap may also contain references to other TimeMaps and TimeGates.

A Memento aggregator is a software implementation of Memento that takes a URI as the parameter, queries multiple supported archives, combines and temporally orders the returned mementos, and returns this list as a TimeMap to the user. We used the MemGator \cite{memgator} implementation of a Memento aggregator in collecting data for our analysis. MemGator provides its own heuristic for determining and reporting the number of mementos present in an aggregated TimeMap using the non-standard \texttt{X-Memento-Count} HTTP header. Despite this, we used the contents of the aggregated TimeMap returned from MemGator instead of this header as the basis for further investigating the number of mementos present.

\subsection{URI Canonicalization}
\label{sec:canonicalization}

URI canonicalization associates differently formatted URIs \cite{rfc6596}. For example, \hyperref[]{http://example.com} might be associated with:
\begin{itemize}
\item \hyperref[]{http://www.example.com}
\item \hyperref[]{https://www.example.com}
\item \hyperref[]{http://example.com/}
\item \hyperref[]{http://example.com/index.html}
\item \href{http://example.com/\string#articles}{http://example.com/\#articles} 
\end{itemize}

Canonicalization allows after-the-fact clustering of URIs that likely reference the same resource. As URI schemes from a Web site change over time, canonicalization is critical for retaining a cohesive, comprehensive listing of the mementos available for a Web page. Internet Archive's Wayback CDX Server API\footnote{\hyperref[]{https://github.com/internetarchive/wayback/tree/master/wayback-cdx-server}} and endpoint\footnote{Example access at \hyperref[]{http://web.archive.org/cdx/search/cdx?url=example.com}} is one of multiple endpoints\hreffootnote{https://archive.org/help/wayback_api.php}{https://archive.org/help/wayback_api.php} that provides access to the indexes of the archive's holdings. A partial example (that corresponds with the TimeMap shown in Figure~\ref{fig:tm_link}) of the data returned from Internet Archive's CDX server is shown in Figure~\ref{fig:cdx}. As an alternative to their Memento endpoint \cite{wsdl-ia-memento}, the CDX endpoint provides the HTTP status code of the capture as well as Sort-friendly URL Reordering Transformed (SURT) URIs. Part of the SURT-generation process involves canonicalizing the \URIRns.  A canonicalized URI is present in the first space-delimited field (Figure~\ref{fig:cdx}) where the ``www'' subdomain is not present despite being part of the URI in the query parameter. The non-canonicalized \URIR attributed to the record in the CDX is available as the third field in the CDX record. Figure~\ref{fig:cdx} shows the \URIR variations including no subdomain, the ``www'' subdomain, with and without a trailing slash, and the explicit inclusion of the port number as all canonicalizing to the same URI in the CDX. 

\begin{figure}
\begin{lstlisting}
com,example)/ 20020120142510 http://example.com:80/ text/html 200 HT2DYGA5UKZCPBSFVCV3JOBXGW2G5UUA 1792
com,example)/ 20020804094019 http://www.example.com:80/ text/html 200 UY3I2DT2AMWAY6DECFCFYMT5ZOTFHUCH 457
com,example)/ 20160728014649 http://www.example.com/ unk 302 3I42H3S6NNFQ2MSVX7XZKYAYSCX5QBYJ 339
com,example)/ 20160728114745 http://www.example.com unk 302 3I42H3S6NNFQ2MSVX7XZKYAYSCX5QBYJ 340
com,example)/ 20160728123024 http://example.com/ text/html 200 ASIFPQKKLDWATFDIO1OJJ3NSK34KLLMN 577
\end{lstlisting}
\caption{A CDX response returned from Internet Archive's CDX Server. The space-delimited fields are representative of the canonicalized (SURTed) URI, datetime, original URI, MIME type of original document (where applicable), HTTP response code, digest of WARC response record, and length of response record.}
\label{fig:cdx}
\end{figure}

\subsection{Archival Indexing}

A CDX record with a 3XX HTTP status code does not contain the ultimate \URIM that the user will experience when the URI-M is dereferenced. Further, the CDX in Figure~\ref{fig:cdx} is only representative of IA's holdings. The corresponding service of aggregating CDX records in an entity like Memento's concept of combining of TimeMaps from different archives through a Memento aggregator does not exist in standard practice for CDX files. Archives that provide a Memento endpoint are not required and frequently do not expose a CDX endpoint like Internet Archive. This prevents simply referencing all aggregated archives' CDX files for a URI to determine the non-redirecting count of mementos. In this work, we utilize the aggregated holdings of multiple Web archives as well as the CDXJ format \cite{salam-cdxj}, an extension of CDX. MemGator's CDXJ generation is derived from the archives' Memento endpoints, specifically their Link formatted TimeMaps, and transformed into the CDXJ format that allows quicker, more reliable parsing of the datetime that the included URI-Ms represent.

\subsection{Opacity of \URIMsns}
\label{sec:opacity}

\begin{sloppypar}
It is tempting to extract \URIR and Memento-Dateime values directly from \URIMsns.  For example, it is likely that \url{http://web.archive.org/web/20140417054441/http://google.com/} is a memento for \url{http://google.com} appeared at April 17, 2014 at 05:44:41 GMT. However, we cannot be sure until we dereference the URI-M and check its response headers for the values in \texttt{rel="original"} and Memento-Datetime.  While it is unlikely that the IA will deceive us, the \URIM may redirect to another URI-M with a different Memento-Datetime, or in the case of an archived HTTP redirection, the URI-M might end up at an altogether different \URIR \cite{alsum-redirects}.  Furthermore, some archives issue \URIMs without semantics --- for example these URI-Ms are all mementos for \url{google.com} but neither this nor the Memento-Datetime can be ascertained without dereferencing: \url{webcitation.org/query?id=1398456230796350}, \url{archive.is/sz8b9}, and \url{perma.cc/H3YY-BQN5}.  For these reasons, we treat \URIMs as fully opaque \cite{webArchitecture-opacity} and dereference all URI-Ms to extract values for \URIR and Memento-Datetime.
\end{sloppypar}

\section{Related Work}

Bar-Yossef et al. \cite{bar-yossef} introduced the term, ``Soft 404s'' to identify Web pages that report a status code other than HTTP 404 despite the page not existing. Meneses et al. \cite{soft404s} described the process of identifying ``Soft 404s'' based on a signature of the page's contents. In this work we describe ``soft 3XXs'' where content is returned from an archive with a status code of 200 yet the contents of the capture consist of an archived HTTP 3XX redirect. With archives that implement Memento, the Accept-Datetime header instructs the archive to return the originally archived status code (Section~\ref{sec:redirects}).

AlSum et al. \cite{alsum-redirects} analyzed memento redirection patterns relating to HTTP redirects to supply the user with the correct memento when a redirect is encountered in the archives. They introduced the notion of ``URI stability'' to give a quantitative measure of the presence of HTTP 3XX status codes that result when URI-Ms in TimeMaps are dereferenced. 

Rosenthal has discussed Memento aggregator merits and downsides. He cautioned against using TimeMap magnitude for determining the number of URI-Ms available, stating, ``Even if we assume that an archive is correct in announcing that it contains a valid copy of the resource at a particular URL at a particular time, that does not imply that it is willing to satisfy a browser's request for that copy.'' \cite{rosenthal-mementoAggregation}.

\begin{table}
\centering
\begin{tabular}{ l | r }
Archive & Memento count\\\hline
Internet Archive                         & 636,246  \\
Archive-It                               & 62,828   \\
Webcitation                              & 7,551    \\
Stanford Web Archive                     & 4,734    \\
UK National Archives Web Archive         & 1,510    \\
Archive.is                               & 1,301    \\
PRONI Web Archive                        & 173      \\
UK Parliament Web Archive                & 127      \\
\hline
Total & 714,470\\
\end{tabular}
\caption{Distribution of mementos for google.com for a collection of archives defined by a locally deployed Memento aggregator.}
\label{tab:google}
\end{table}

\section{Data Collection}
\label{sec:dataCollection}
To analyze the degree to which archival identifiers result in redirects, we needed to acquire the HTTP response headers for all URI-Ms accumulated from multiple Web archives for a URI-R. The concept of a Memento aggregator allows us to accomplish this task, albeit parsing the standardized Link formatted resulting TimeMaps is potentially prone to error.

We deployed a local instance of MemGator\footnote{\hyperref[]{http://github.com/oduwsdl/memgator}} version 1.0-RC4 configured to query the archives listed in Table~\ref{tab:google}. The MemGator instance was initialized with 25 minutes as the value for the ``restimeout'' (response timeout for each archive) and ``hdrtimeout'' (header timeout for each archive) parameters. Declaring these timeout values ensured that the server portion of data collection was not prematurely returned because of network latency with communication to the considered archives. Our client script that queried the local MemGator instance was also setup to access this instance with equally large timeout values. 

We leveraged MemGator's CDXJ \cite{salam-cdxj} interface (example output in Figure~\ref{fig:tm_cdxj}) for simple datetime extraction, structured JSON-formatted metadata of each memento's attributes, and more human readable output compared to the conventional Link (Figure~\ref{fig:tm_link}) or JSON formatted TimeMaps. Collection was run on a late 2013 MacBook Pro running OS X version 10.11.4 with a 2.4 GHz Intel i5 processor, 8 GB of RAM, and a 250 GB SSD disk. Data was collected mid-May, 2016. We performed an initial analysis of the mementos contained within the TimeMap without dereferencing any mementos. The client script to query the local MemGator was created in Python 2.7.10 using the ``requests'' library\footnote{\hyperref[]{http://python-requests.org}} and the built-in JSON parser.

\begin{figure}
\begin{lstlisting}[basicstyle=\footnotesize]
@meta {"original_uri": "http://example.com"}
@meta {"timegate_uri": "http://localhost:1208/timegate/http://example.com"}
@meta {"timemap_uri": {
 "link_format": "http://localhost:1208/timemap/link/http://example.com", 
 "json_format": "http://localhost:1208/timemap/json/http://example.com",
 "cdxj_format": "http://localhost:1208/timemap/cdxj/http://example.com"
 }
...
20090418233448 {"uri": "http://web.archive.org/web/20090418233448/http://www.example.com/", "rel": "memento", "datetime": "Sat, 18 Apr 2009 23:34:48 GMT"}
20090421223547 {"uri": "http://wayback.vefsafn.is/wayback/20090421223547/http://www.example.com/", "rel": "memento", "datetime": "Tue, 21 Apr 2009 22:35:47 GMT"}
20090421231335 {"uri": "http://webarchive.loc.gov/all/20090421231335/http://www.example.com/", "rel": "memento", "datetime": "Tue, 21 Apr 2009 23:13:35 GMT"}
...
\end{lstlisting}
\caption{A partial CDXJ formatted TimeMap returned from a local instance of MemGator containing URI-Ms from multiple archives.}
\label{fig:tm_cdxj}
\end{figure}

\section{Analysis based on TimeMaps}
\label{sec:initialAnalysis}

We obtained a TimeMap for google.com from our locally deployed Memento aggregator (MemGator instance) containing 714,470 URI-Ms from 8 different Memento-compliant archives. Table~\ref{tab:google} shows the distribution of the mementos using a simple URI-based association algorithm to attribute URI-Ms to an archive. 89.1\% of the URI-Ms returned were from Internet Archive.

\subsection{Variation in Scheme}
Two schemes \cite{rfc7230} are used for the URI-Rs contained within the URI-Ms returned: HTTP and HTTPS. Table~\ref{tab:scheme} shows the breakdown of the URI-Rs contained in the URI-Ms based on the TimeMap, inclusive of the inferred \URIRs whose scheme could not be determine solely from the \URIMns. As discussed in Section~\ref{sec:opacity}, the more accurate method to attribute a \URIR to a \URIM is to obtain the memento's ``original'' Link header values but this section focuses on analyzing the contents of the TimeMap without requesting the \URIMsns.

From the URI-Rs that could be extracted, 86.2\% used the HTTP scheme.
Table \ref{tab:google2} shows the \URIRns-based memento count for each scheme using a substring-based grouping approach similar to that used for Table~\ref{tab:google}. The TimeMap contained canonicalized variants of the URI-Rs embedded as substrings within the URI-Ms with our query for the CDXJ-formatted TimeMap. Our query supplied the \URIR variant containing the HTTP scheme, www sub-domain, and no trailing characters combination (i.e., \hyperref[]{http://www.google.com}) via the query to MemGator\footnote{\hyperref[]{http://localhost:1208/timemap/cdxj/http://www.google.com}}.

\begin{table}
\centering
\begin{tabular}{ r  r }
Scheme & URI-R count \\\hline
http & 609,274 \\
https & 97,645 \\
unknown & 7,551 \\ \hline
 & 714,470
\end{tabular}
\caption{Scheme distribution among the URI-Rs within the mementos in the TimeMap for google.com.}
\label{tab:scheme}
\end{table}

\begin{figure}
\centering
\includegraphics[width=1.0\linewidth]{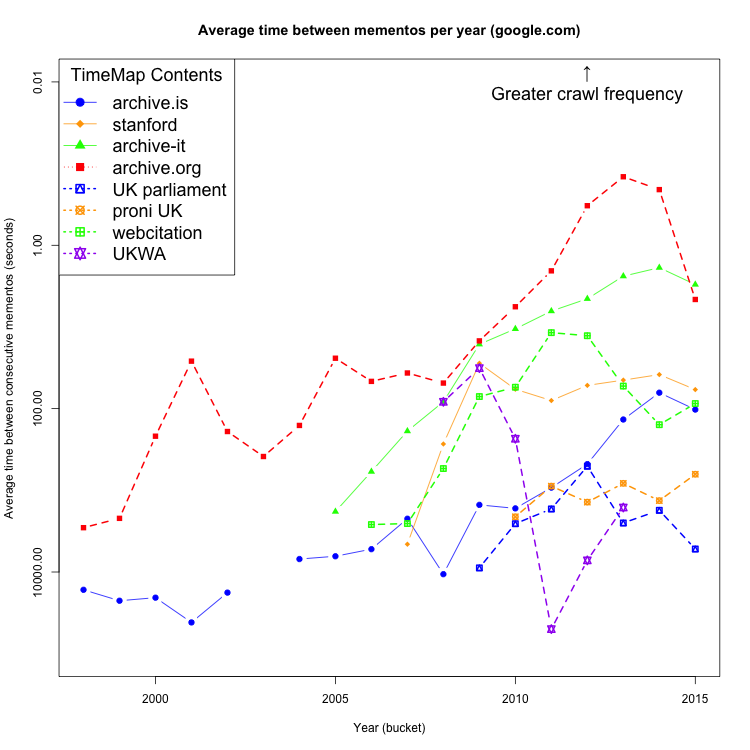}
\caption{The average time between consecutive mementos has decreased with time. This plot shows a year-based bucketing of the difference in time between adjacent mementos from different archives for google.com. }
\label{fig:avgTime}
\end{figure}

\subsection{Grouping by Year}
We separated the URI-Ms as reported by the TimeMap into year-based buckets using the ``datetime'' attribute for each \URIM (and not the embedded 14-digit date stamp per Section~\ref{sec:opacity}) as well as by-archive for google.com. We calculated the average time between URI-Ms within a year-based bucket to show that the velocity of capturing google.com is generally increasing in time (Figure~\ref{fig:avgTime}). The quantity of captures from Internet Archive for google.com in 2015 was significantly lower than the trend would indicate. Table~\ref{tab:tcOverTM} indicates this dramatic drop in google.com captures from both the IA CDX endpoint and from the Memento endpoint. Also because data collection occurred in May 2016, the partial year data points for 2016 are on-par with the trend of years prior to 2015.

\begin{table}
\centering
\begin{tabular}{ r  l  r }
Scheme & Format & \URIR count \\ \hline
http & http://www.google.com & 541,160 \\
     & http://google.com     & 67,811 \\
     & http://\textit{other}.google.com     & 303
\\ \hline
https & https://www.google.com & 96,853 \\
      & https://google.com & 792 \\
      & https://\textit{other}.google.com     & 0
\\ \hline
\end{tabular}
\caption{Count of URI-Rs contained within URI-Ms for google.com.} 
\label{tab:google2}
\end{table}

\begin{table}
\centering
\begin{tabular}{ r r r r }
year & M$_{TM}$ & M$_{RC}$ & $DI$\\\hline
1998 &        4 &        4 & $\infty$ \\
1999 &       19 &       19 & $\infty$ \\
2000 &      132 &       87 & 1.933 \\
2001 &    1,185 &      579 & 0.955 \\
2002 &      176 &      137 & 3.513 \\
2003 &       75 &       55 & 2.750 \\
2004 &      197 &      143 & 2.648 \\
2005 &    1,236 &      414 & 0.504 \\
2006 &      735 &      483 & 1.917 \\
2007 &    1,055 &      842 & 3.953 \\
2008 &    1,376 &      894 & 1.855 \\
2009 &    6,074 &    4,335 & 2.493 \\
2010 &    9,326 &    6,530 & 2.335 \\
2011 &   20,634 &    9,279 & 0.817 \\
2012 &  102,533 &   16,240 & 0.188 \\
2013 &  228,405 &   25,203 & 0.124 \\
2014 &  164,865 &   22,738 & 0.160 \\
2015 &   17,978 &   11,286 & 1.686 \\
2016 &  139,520 &    5,805 & 0.043 \\
\hline
\end{tabular}
\caption{Google over time, bucketed by year, based on IA mementos extracted from the MemGator CDXJ TimeMap. M$_{TM}$ is the memento count based solely on the data in the TimeMap, M$_{RC}$ is the count based on exclusion of redirects when dereferenced, and $DI$ is the ratio of non-redirecting mementos to redirecting mementos, per Section~\ref{sec:canonicalizationPatterns}.}.
\label{tab:tcOverTM}
\end{table}

\subsection{``TimeMap'' from CDX Server}
We also obtained the CDX for google.com from Internet Archive (IA). We compared the HTTP response codes we received when dereferencing the IA URI-Ms from the CDXJ TimeMap (Section~\ref{sec:dataCollection}) with the response codes explicitly provided in the CDX listing IA returned. Per Section~\ref{sec:canonicalization}, a CDX endpoint is not available as a user-accessible endpoint from most Web archives. We used the available endpoint at IA as a sanity check for correctness of the data obtained when the URI-Ms in a TimeMap are dereferenced. The intention of analyzing the TimeMap and not simply deferring to the CDX Server, despite the majority of mementos in the aggregated TimeMap being from IA, is to extrapolate the dereferencing strategy to other Memento-compliant Web archives.

\section{Analysis based on Mementos}
\label{sec:mementoBasedAnalysis}
In Section~\ref{sec:initialAnalysis} we analyzed the archival presence of a URI based solely on the TimeMap supplied by a local aggregator when querying the aggregator with one canonicalized variant of the URI-R. In this section, we dereference the URI-Ms in the TimeMap for further analysis. Equations~\ref{eq:naive} and \ref{eq:naive2} set the basis for counting mementos in a TimeMap by merely counting the entries where the ``rel'' attribute in a TimeMap contains a value of ``memento''. For example, Figure~\ref{fig:tm_link} shows a Link formatted TimeMap where each highlighted entry containing a ``memento'' rel value (with the potential additional inclusion of other values like ``last'' and ``first'') increments the count according to Equation~\ref{eq:naive2}.

\begin{equation}\label{eq:naive}
M = \text{URI-M} \subset \text{TimeMap}, \text{if}\ ``memento" \subset \text{valuesOf(}rel\text{)}\\
\end{equation}
\begin{equation}\label{eq:naive2}
|TM|_{rel} = \sum_{m=1}^{len(M)} 1
\end{equation}

\subsection{Redirects in Mementos}
\label{sec:redirects}
The number of non-redirecting (non-3XX) mementos in a TimeMap cannot be counted with the TimeMap data alone. When URI-Ms are dereferenced, they need not contain an entity body but may consist only of an archived HTTP response, which might not be a 200. This occurs in cases where the live Web site returned an HTTP 302 redirect, among other circumstances. This redirect was captured, retained, and is replayed by the archives. When replayed, the datetime originally requested for a URI-R, as is often the case, will be different than the datetime of the memento ultimately served to the user. This ``archived 302'' is different from a 3XX code returned from an archive that is not representative of an archival capture; for example, when a datetime for a URI is requested where no capture for the URI is contained within the archive's holdings. 

\begin{sloppypar}
Assuming the TimeMap in Figure~\ref{fig:tm_link} is wholly inclusive of all of the mementos contained by Internet Archive for example.com, requesting the \mbox{URI-M} \hyperref[]{http://web.archive.org/web/20160728114743/http://www.example.com} (two seconds before a listed \URIM neglecting datetime semantics per Section~\ref{sec:bg}) will result in a 302 from the archive pointing to the nearest capture. This behavior is a function of the archive, is not mandatory behavior to exhibit, and is functionality independent of the Memento protocol. Were there a capture at the former datetime where the archival crawler experienced a 302 from the live Web at the time, the TimeMap would contain the \URIM \hyperref[]{http://web.archive.org/web/20160728114743/http://www.example.com} with a rel value of ``memento'' indistinguishable from the memento entry at \hyperref[]{http://web.archive.org/web/20160728114743/http://www.example.com} regardless of the status code that occurs from the archive when each \URIM is dereferenced.
\end{sloppypar}

\subsection{Direct and Indirect Mementos}
\label{sec:userAgent}

Users interacting with an archive via a Web browser will not directly experience intermediary HTTP transactions (the user agent automatically redirects the user to a non-3XX status), we introduce the term URI-M$_{D}$ (for ``direct'') to indicate a \URIM in a TimeMap that does not require any intermediary transaction to resolve. Thus, a URI-M$_{D}$ is a case of a \URIM where the \URIM originally requested by the user is the identifier for the ultimate memento served. URI-Ms in a TimeMap that exhibit the behavior where a HTTP 3XX class code is replayed and the datetime differs from that requested by the user are indicated with URI-M$_{I}$ (for ``indirect'').

In Equation~\ref{eq:rm3XX} we filter $|TM|_{rel}$ from Equation~\ref{eq:naive2} to exclude mementos that resolve to HTTP 3XX status codes. $|TM|_D$ represents the count of mementos that result in non-3XX statuses based on the URI-Ms in a TimeMap. Section 2.2.4 of the Memento RFC \cite{rfc7089} states that a link with a datetime attribute must match the value of the Memento-Datetime header when the link is dereferenced.

\begin{equation}\label{eq:rm3XX}
|TM|_D = \sum_{m=1}^{len(M)} 
\begin{cases}
0\ & 300 \geq httpStatus(m)  < 400,\\
1\ & \text{otherwise.}
\end{cases}
\end{equation}

\begin{equation}\label{eq:just3XX}
|TM|_I = |TM|_{rel} - |TM|_D
\end{equation}

We quantify the ratio of mementos with non-redirecting HTTP status codes (Equation~\ref{eq:rm3XX}) to those with redirects (Equation~\ref{eq:just3XX}) in Equation~\ref{eq:rat} as $DI$.

\begin{equation}\label{eq:rat}
DI =
\begin{cases}
\frac{|TM|_D}{|TM|_I}\ & |TM|_I > 0,\\
\infty & \text{otherwise.}
\end{cases}
\end{equation}

\begin{figure}
\centering
\includegraphics[width=1.0\linewidth]{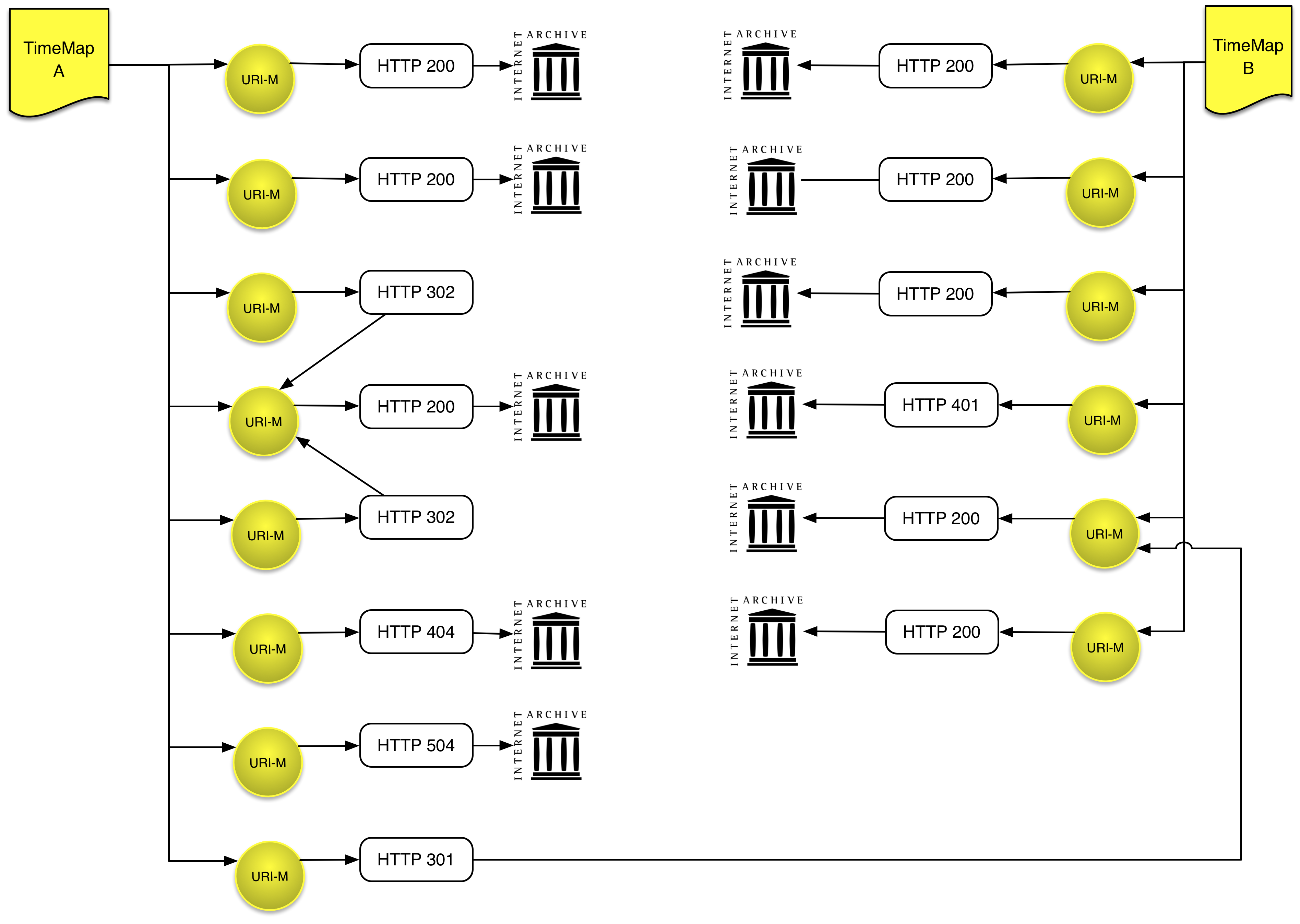}
\caption{
Dereferencing URI-Ms in a TimeMap may produce a variety of HTTP status code, some of which are redirects both to other URI-Ms within the TimeMap and URI-Ms not included in the initial TimeMap. Counting the number of mementos without dereferencing URI-Ms is therefore problematic.}
\label{fig:balls}
\end{figure}

\label{sec:scenario}
As an example, Figure~\ref{fig:balls} contains 11 URI-Ms that result in non-redirecting archived HTTP status codes when dereferenced, inclusive of eight 200 codes, two 4XX codes, and one archived 504. TimeMap A represents a domain for an organization (for example) that is acquired by another organization, whose domain is represented by TimeMap B. At the point of acquisition, TimeMap A redirects to TimeMap B, as represented by the HTTP 301. At two points prior to acquisition, an archival crawler attempted to capture the URI-R for TimeMap A but received a server-side redirect, which is reflected in the preserved HTTP 302 responses. As the acquisition proceeded, the URI-R may have been deleted (the 404 in TimeMap A) and the server misconfigured in the transition (HTTP 504). An intermittent HTTP 401 (Unauthorized) error is also experienced in the URI-R for TimeMap B as the servers are reconfigured to accept the additional traffic from the acquisition. 

Three \URIMs resulted in 3XX redirects when dereferenced. Using Equation~\ref{eq:rat}, $DI = 11/3 \approx 3.7$. Sparsely archived URIs will often contain a list of URI-Ms where all result in an HTTP 200 status code when dereferenced, which would result in $DI$ being undefined. It is far less likely that all URI-Ms in a TimeMap return in an HTTP redirect when dereferenced.

During the data acquisition process (Section~\ref{sec:dataCollection}) we experienced intermittent HTTP 5XX status codes \cite{rfc7231} in responses from Internet Archive, namely HTTP 503 (Service Unavailable) and 504 (Gateway Timeout). In much of the same way that sending an Accept-Datetime header causes a ``soft'' HTTP status code to ``harden'', we repeated the request via curl\footnote{\url{https://curl.haxx.se/}} with the inclusion of an Accept-Datetime HTTP header \cite{rfc7089}. This additional step caused no change in the subsequently returned results compared to the originally results response. Repeating the collection procedure for select URI-Ms in the future remedied this issue, allowing us to attribute the error to the archive and not the archive returning a capture of an archived 5XX. If the response instead indicated that the returned 5XX status codes was representative of the state of the \URIR at the respective time (through providing a Memento-Datetime response header) and not an intermittent result attributable to the archive, the \URIM would signify an increment in the Equation~\ref{eq:rm3XX} summation.
 
\subsection{Canonicalization Patterns}
\label{sec:canonicalizationPatterns}
In observing the mementos for \hyperref[]{http://www.google.com}, we encountered 3 canonicalization patterns for URI-Ms that surface those that are URI-M$_{D}$ versus those that are URI-M$_{I}$. We define $M_{TM}$ to be the memento count for a TimeMap when using only the data contained in the TimeMap without dereferencing mementos (Section~\ref{sec:initialAnalysis}). We define the {\em representative count} ($M_{RC}$) of the number of mementos present in a TimeMap to be the number of URI-M$_{D}$ where $M_{RC} \leq M_{TM}$. These canonicalization patterns observed are Inter-scheme, Slash-added, and subdomain redirect patterns, described in this section.

\def\createlinenumber#1#2{
    \edef\thelstnumber{%
        \unexpanded{%
            \ifnum#1=\value{lstnumber}\relax
              #2%
            \fi}%
        \ifx\thelstnumber\relax\else
        \expandafter\unexpanded\expandafter{\thelstnumber}%
        \fi
    }
}
\let\thelstnumber\relax
\createlinenumber{3}{$M_1$}
\createlinenumber{5}{$M_2$}
\createlinenumber{6}{$M_3$}
\createlinenumber{8}{$M_4$}
\createlinenumber{9}{$M_5$}
\createlinenumber{11}{$M_6$}
\createlinenumber{12}{$M_7$}

\begin{figure*}[t]
\begin{lstlisting}[language=bash, basicstyle=\footnotesize, numbers=left,stepnumber=1, showstringspaces=false,tabsize=1,breaklines=true,breakatwhitespace=false]
<http://google.com>; rel="original",
...
<http://web.archive.org/web/20011124163711/http://www2.google.com/>; rel="memento"; datetime="Sat, 24 Nov 2001 16:37:11 GMT", 
...
<http://web.archive.org/web/20130101000813/http://www.google.com/>; rel="memento"; datetime="Tue, 01 Jan 2013 00:08:13 GMT", 
<http://web.archive.org/web/20130101003310/https://www.google.com/>; rel="memento"; datetime="Tue, 01 Jan 2013 00:33:10 GMT", 
...
<http://web.archive.org/web/20140425221431/http://www.google.com>; rel="memento"; datetime="Fri, 25 Apr 2014 22:14:31 GMT",
<http://web.archive.org/web/20140425221433/https://www.google.com/>; rel="memento"; datetime="Fri, 25 Apr 2014 22:14:33 GMT", 
...
<http://web.archive.org/web/20160519223823/http://www.google.com/>; rel="memento"; datetime="Thu, 19 May 2016 22:38:23 GMT", 
<http://web.archive.org/web/20160520165954/http://google.com/>; rel="memento"; datetime="Fri, 20 May 2016 16:59:54 GMT",
...
\end{lstlisting}
\caption{A partial TimeMap in Link format for google.com with annotations highlighting various URI-Ms, discussed in Section~\ref{sec:canonicalizationPatterns}.}
\label{fig:tm_link_Ms}
\end{figure*}

\subsubsection{Inter-scheme \URIM Redirect}
\label{sec:interScheme}

As adoption of the secure HTTPS scheme over HTTP becomes more prevalent on the live Web \cite{freedompress, httpsBreaking}, the trend becomes apparent in the archives through canonicalizing the HTTP and HTTPS site to be one in the same. For example, observe two mementos from 2013 from the TimeMap for \hyperref[]{google.com} (Figure~\ref{fig:tm_link_Ms}). The status code returned for $M_2$ is 200 with no HTTP location header present (an example of a URI-M$_{D}$). However, the status code returned for $M_3$ is an HTTP 302 with an HTTP location response header of /web/20130101000813/http://www.google.com/, i.e., $M_3$ redirects to $M_2$ when dereferenced. Thus, $M_3$ is a URI-M$_{I}$. Were the naive but often applied Equation~\ref{eq:naive2} used for determining how many mementos are represented by the URI-Ms $M_2$ and $M_3$, both would be included while dereferencing each \URIM would result in a count of only a single memento. This highlights an important distinction and discrepancy between the number of identifiers (URI-Ms) and the number of representations (mementos).

\subsubsection{Slash-added \URIM Redirect}
\label{sec:slashAdded}

Queries for the URI \hyperref[]{http://www.google.com} are sometimes redirected to the same URI with an appended slash. For example, the two mementos $M_4$ and $M_5$ both exist in a TimeMap (Figure~\ref{fig:tm_link_Ms}).

When dereferenced, $M_4$ returns a 302 with a location header pointing to $M_5$, captured two seconds later based solely on the embedded datetime. Both URI-Ms are reported by the TimeMap while only the latter contains an entity body when dereferenced.

\begin{table}
\centering
\begin{tabular}{ r | r }
Time Gap Bucket										&URI-M count\\\hline
0 seconds											&24,541\\
1 second										    &34,577\\
2 seconds											&26,153\\
3 seconds											&62,526\\
4 seconds											&46,738\\
5 seconds											&14,215\\
6 seconds											&12,213\\
7 seconds											&9,748\\
8 seconds											&7,431\\
9 seconds											&6,610\\
$>$ 9 seconds, $\leq$ 1 minute		&101,868\\
$>$ 1 minute, $\leq$ 1 hour              &247,192\\
$>$ 1 hour, $\leq$ 1 day                   &37,399\\
$>$ 1 day                                      &5,034\\ \hline
\end{tabular}
\caption{A range of time differences exists between adjacent captures of a \URIRns. This table represents the instances of these differences between adjacent URI-Ms from the TimeMap for google.com reported by Internet Archive.}
\label{tab:google4}
\end{table}

\subsubsection{Subdomain \URIM Redirect}
\label{sec:subdomainSwitch}

It is also useful to observe canonicalization that does not result in a redirect. Google has used a variety of subdomains of the sort containing the literal ``www'' followed by a digit over the years, as with $M_1$ (Figure~\ref{fig:tm_link_Ms}). Accessing this memento (dereferencing $M_1$) results in an HTTP 200 status code. The other, much more common subdomain of www, as with $M_6$ returns an HTTP 302 redirected to $M_7$, both present in the TimeMap. Table~\ref{tab:datatatata} shows the magnitude of redirects for google.com based on the URI-R$_{orig}$ scheme, URI-R$_{orig}$ subdomain, URI-R$_{dest}$ scheme, and URI-R$_{orig}$ subdomain. The breakdown in Table~\ref{tab:datatatata} was also generated for comparison to the \URIRs vimeo.com and wikipedia.org in Tables~\ref{tab:datatatata_vimeo} and ~\ref{tab:datatatata_wikipedia}, respectively. 

\subsubsection{Analysis}
\label{sec:analysis}

\begin{figure}
\centering
\includegraphics[width=1.0\linewidth]{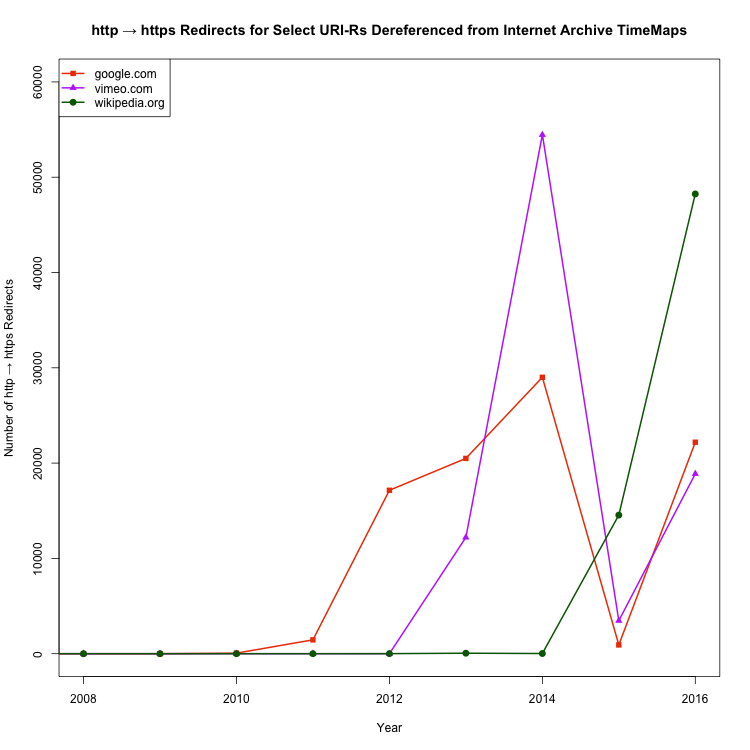}
\caption{Inter-scheme redirects for google.com, vimeo.com, and wikipedia.org from the mementos in the TimeMap from MemGator for IA that result in a 3XX. }
\label{fig:interScheme}
\end{figure}

Juxtaposing the proportion exhibited for each of the four permutations of scheme transitions (HTTP-to-HTTP, HTTP-to-HTTPS, etc.) from vimeo.com and wikipedia.org as compared to google.com, the inter-scheme transition seems more common with the former pair while the bulk of the results for google.com reside in redirects that retain both the HTTP scheme in the \URIR but also the www subdomain. Focusing specifically on the HTTP-to-HTTPS inter-scheme transition, Figure~\ref{fig:interScheme} serves as an interesting cross-section of Table~\ref{tab:datatatata} broken down by time. Disregarding the anomalous captures from Internet Archive in 2015, the overall trend of inter-scheme redirects is leaning toward more HTTP-to-HTTPS than HTTPS-to-HTTP as the secure scheme is adopted by more sites on the live Web. Disregarding 2015, the HTTP-to-HTTPS redirects for vimeo.com appear to be monotonically increasing with normalization for the partial year results for 2016 (collection was performed in May of that year). Figure~\ref{fig:interScheme} also shows a steeper quantity of captures containing these redirects for wikipedia.org with few captures of redirects exhibiting this inter-scheme permutation prior to 2014. The rapid increase in each site may temporally correspond with the adoption of the HTTPS scheme by the live Web site, thereby forwarding all traffic accessing the HTTP version of the site using an HTTP 3XX response.

An additional nuance to account for the large quantity of redirects from HTTP \URIMs to HTTP \URIMs for google.com can be observed by the large quantity of ``revisit'' entries in IA's CDX results for google.com. A revisit entry occurs when an archival crawler is returned content that is identical to a previous capture, often attributed using a hashing scheme on the live page's content. If an archive reports revisit records as an HTTP redirect based on the CDX listing, and this redirect is propagated to the archive's Memento endpoint thus producing a unique URI-M, the $DI$'s value for the \URIR decreases. Requesting the \URIM using the Accept-Datetime HTTP header then observing the Memento-Datetime response header's presence often reveals this nuance, but by relying on the TimeData data without requesting each \URIM, the $DI$ for the \URIR is unknown.

\begin{table}
\centering
\begin{tabular}{ c | r }
Year	&Memento count\\\hline
2001 & 68 \\
2005 & 391 \\
2006 & 8 \\
2007 & 81 \\
2008 & 62 \\
2009 & 153 \\
2010 & 124 \\
2011 & 616 \\
2012 & 5,564 \\
2013 & 25,914 \\
2014 & 40,819 \\
2015 & 1,367 \\
2016 & 10,104 \\
\end{tabular}
\caption{Google memento analysis group by memento year where the time between two mementos is less than or equal to 2 seconds.}
\label{tab:lessThan2}
\end{table}

\subsection{Inter-Memento Temporality}
We measured the time between each pair of consecutive mementos, shown in Table \ref{tab:google4}. We found that 38.4\% had a time gap less than 9 seconds, indicating (in some cases) that a redirect would occur when the \URIM is dereferenced. Using a yearly bucketing scheme (Table~\ref{tab:allThePlots}), we plotted the difference in time between adjacent mementos from IA based on the scheme in the URI-M$_{I}$ and the ultimate URI-M that results when the URI-M$_{I}$ is dereferenced. For each log-log plot, shown with more details in the Appendix, a point's quadrant positioning is indicative of the quantity of mementos with a seconds-level granularity of time. For example, a point in the top-left quadrant of a plot indicates that there are many temporally consecutive memento pairs with a very small time difference between them. Top right would indicate many pairs with a large time difference between them; bottom right: few memento pairs with a large time difference; bottom left: few memento pairs with a small time difference. Many more points in the left half of a plot than the right indicates much less time between captures, i.e., the capture frequency was higher that year. More points being in the right half of the plot indicates that more time passes between consecutive captures. The trend for google.com excluding 2015 shows fewer pairs with a small time difference (more points in the bottom right) as time goes on for all redirect patterns other than HTTP-to-HTTP.

\subsection{Temporal Closeness as an Indicator of Redirection}
In only brief examination of the TimeMap, some temporally consecutive URI-Ms appeared in ``pairs'' where a second URI-M exists from the same archive within seconds of the previous. Table~\ref{tab:lessThan2} lists the by-year breakdown filtering to only include the URI-Ms where the time between the two is less than two seconds. The trend generally increases with time. This plot can also be cross-referenced with Table~\ref{tab:datatatata}, which shows the overall inter-scheme redirect breakdown totals independent of time with the additional subdomain granularity.

Table~\ref{tab:lessThan2} also shows a peak in 2014 at 40,819 pairs where the redirect is less than or equal to two seconds apart from the URI-R$_{orig}$ to the URI-R$_{dest}$. Given the quantity of inter-scheme redirects in Figure~\ref{fig:interScheme} for 2014 totaling around 30,000 as the sum and Table~\ref{tab:datatatata} showing a significantly larger number of same scheme redirects (e.g., 490,836 just for HTTP-to-HTTP both with the www subdomain), many redirects over the archived history of google.com can be attributed to something other than a scheme switch. The large number of aforementioned identical scheme and subdomain redirects indicates patterned responses like slash-added (Section~\ref{sec:slashAdded}) rather than scheme (Section~\ref{sec:interScheme}) or subdomain switch (Section~\ref{sec:subdomainSwitch}).

\begin{table}
\centering

\begin{subtable}{\textwidth}
\centering
\begin{tabular}{| r | r | r | r | r | r | r | r | r | r |}
\hline

\diagbox{URI-R$_{orig}$ scheme}{URI-R$_{dest}$ scheme} & 
& \multicolumn{3}{|c|}{http} & \multicolumn{3}{|c|}{https} \\  \cline{3-8}
\multirow{3}{*}{http} & & none & www & other & none & www & other \\ \hline
& none & 1,279 & 68,837 & 55 & \ooorange{12} & \ooorange{20,825} & \ooorange{27} \\
& www & 8,934 & 490,836 & 204 & \ooorange{32} & \ooorange{77,610} & \ooorange{16} \\
& other & 0 & 224 & 22 & \ooorange{0} & \ooorange{26} & \ooorange{2} \\ \hline

\multirow{3}{*}{https} & none & 14 & 731 & 0 & 0 & 296 & 1 \\
& www & 1,117 & 72,874 & 27 & 15 & 18,525 & 2,101 \\
& other & 0 & 0 & 0 & 0 & 0 & 0 \\ \hline

\end{tabular}
\caption{Scheme and subdomain for redirects when dereferencing URI-Ms for google.com.}
\label{tab:datatatata}
\end{subtable}

\begin{subtable}{\textwidth}
\centering
\begin{tabular}{| r | r | r | r | r | r | r | r | r | r |}
\hline

\diagbox{URI-R$_{orig}$ scheme}{URI-R$_{dest}$ scheme} & 
& \multicolumn{3}{|c|}{http} & \multicolumn{3}{|c|}{https} \\  \cline{3-8}
\multirow{3}{*}{http} & & none & www & other & none & www & other \\ \hline
& none & 1,642 & 104 & 0 & \greeeeen{82,637} & \greeeeen{0} & \greeeeen{0} \\
& www & 1,273 & 50 & 0 & \greeeeen{6,355} & \greeeeen{0} & \greeeeen{0} \\
& other & 0 & 0 & 0 & \greeeeen{0} & \greeeeen{0} & \greeeeen{0} \\ \hline

\multirow{3}{*}{https} & none & 315 & 6 & 0 & 35,293 & 1 & 0 \\
& www & 10 & 0 & 0 & 1,149 & 0 & 0 \\
& other & 0 & 0 & 0 & 0 & 0 & 0 \\ \hline

\end{tabular}
\caption{Scheme and subdomain for redirects when dereferencing URI-Ms for vimeo.com.} 
\label{tab:datatatata_vimeo}
\end{subtable}

\begin{subtable}{\textwidth}
\centering
\begin{tabular}{| r | r | r | r | r | r | r | r | r | r |}
\hline

\diagbox{URI-R$_{orig}$ scheme}{URI-R$_{dest}$ scheme} & 
& \multicolumn{3}{|c|}{http} & \multicolumn{3}{|c|}{https} \\  \cline{3-8}
\multirow{3}{*}{http} & & none & www & other & none & www & other \\ \hline
& none & 91 & 10,575 & 0 &  \brown{0} &  \brown{4,140} & \brown{0} \\
& www & 110 & 5,099 & 0 &  \brown{1} &  \brown{44,104} & \brown{0} \\
& other & 0 & 0 & 0 & 0 &  \brown{0} &  \brown{0} \\ \hline

\multirow{3}{*}{https} & none & 1 & 46 & 0 & 1 & 804 & 0 \\
& www & 14 & 1,014 & 0 & 0 & 5,602 & 0 \\
& other & 0 & 0 & 0 & 0 & 0 & 0 \\ \hline
\end{tabular}
\caption{Scheme and subdomain for redirects when dereferencing URI-Ms for wikipedia.org.} 
\label{tab:datatatata_wikipedia}
\end{subtable}

\caption{When URI-Ms for three select domains (google.com, vimeo.com, and wikipedia.org), are dereferenced and produce an HTTP redirect, the originally accessed URI-R$_{orig}$ can result in a URI-R$_{dest}$ with a different scheme and subdomain. Cell colors correspond to lines in Figure~\ref{fig:interScheme}. The scheme and subdomain is ``Unknown'' for URI-Ms (like those from webcitation) that obfuscate \URIR with which the \URIM is associated. See Section~\ref{sec:bg}.}
\label{tab:dataTables}

\end{table}

\begin{table}
\centering
  \begin{tabular}{ l | r | r || r | r }
host & \% 3XX & \% 200 & M$_{TM}$ & $DI$ \\\hline
google      & 84.89 & 15.11& 695,525 & 0.178\\
yahoo       & 88.16 & 11.83 & 418,896 & 0.134\\
sourceforge & 73.34 & 26.63 & 31,408 & 0.363\\
instagram   & 67.32 & 32.65 & 55,228 & 0.485\\
vimeo       & 57.04 & 42.94 & 199,262 & 0.752\\
cnn         & 49.97 & 50.01 & 87,148 & 1.001\\
wikipedia   & 44.62 & 55.19 & 25,973 & 1.240\\
whitehouse  & 44.57 & 55.24 & 26,006 & 1.243\\
  \end{tabular}
    \caption{Dereferencing 7 other large Web sites' TimeMaps from Internet Archive produces the above distribution of status codes for each site.}
    \label{tab:sites}
\end{table}

\subsection{Beyond Google}

We then evaluated the applicability of the observations for google.com with other archived Web sites. We dereferenced the TimeMaps of 7 additional large Web sites (Table~\ref{tab:sites}) with a variety of adoption trends of HTTPS and ephemerality as well as 13 home pages of various universities and colleges (Table~\ref{tab:edu}). From this further analysis, we observed how prevalent the trend is as exhibited by google.com with a hypothesis that the relatively static, fundamentally unchanging Google homepage is a reason for the relatively low $DI$.

\begin{figure}
\includegraphics[width=0.32\linewidth]{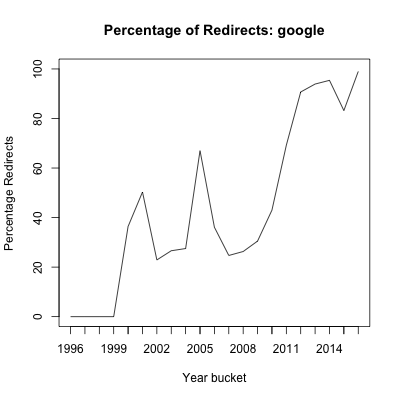}
\includegraphics[width=0.32\linewidth]{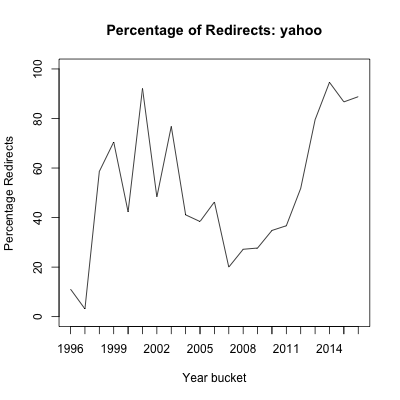}
\includegraphics[width=0.32\linewidth]{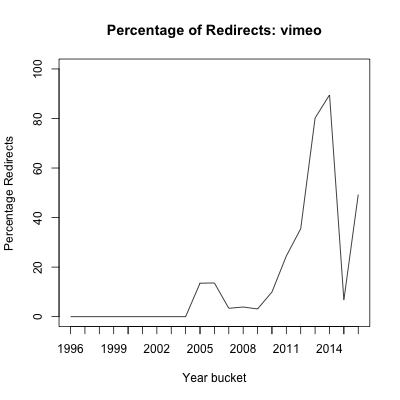}
\includegraphics[width=0.32\linewidth]{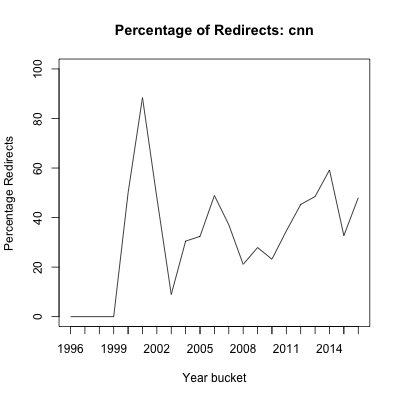}
\includegraphics[width=0.32\linewidth]{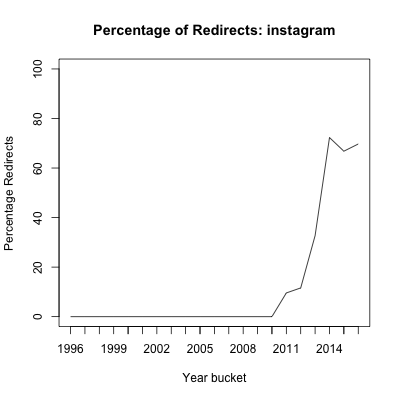}
\includegraphics[width=0.32\linewidth]{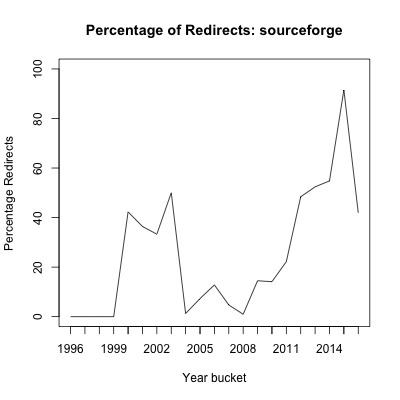}
\includegraphics[width=0.32\linewidth]{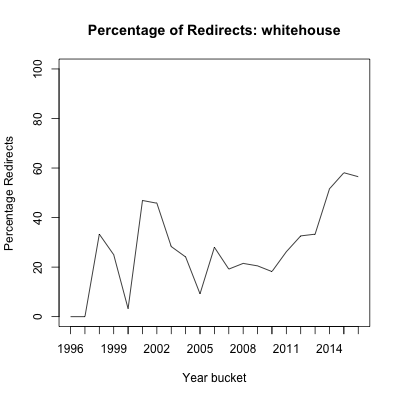}
\includegraphics[width=0.32\linewidth]{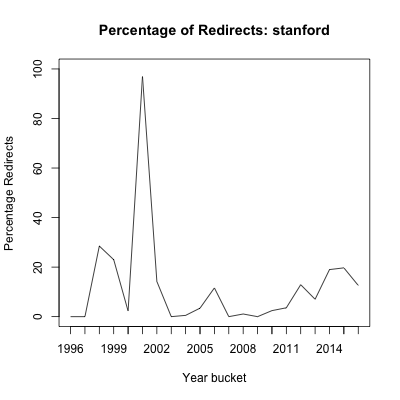}
\includegraphics[width=0.32\linewidth]{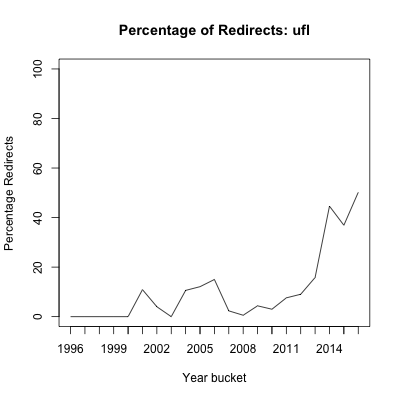}
\caption{Nine URI-Rs from Tables~\ref{tab:sites} and~\ref{tab:edu} exhibit different degrees of redirection over time.}
\label{fig:redirectPlots}
\end{figure}

For the select academic institutions in Table~\ref{tab:edu}, $DI$ is inversely proportion to M$_{TM}$, albeit not strictly as evidenced by ``gatech'' and ``odu''. This pattern does not generally hold in comparison to the large sites in Table~\ref{tab:sites} though the selection of sites for each may contain some inadvertent bias. Figure~\ref{fig:redirectPlots} shows nine plots representing the percentage of redirects over time as determined when all URI-Ms with a rel value in the respective TimeMaps from IA are dereferenced.

\begin{table}
\centering
  \begin{tabular}{ l | r | r || r | r }
host & \% 3XX & \% 200 & M$_{TM}$ & $DI$ \\\hline
stanford    & 62.14 & 37.84 & 19,309 & 0.609\\
princeton   & 60.10 & 39.88 & 9,355 & 0.663\\
columbia    & 48.01 & 51.88 & 9,882 & 1.082\\
harvard     & 33.91 & 65.96 & 7,699 & 1.948\\
caltech     & 33.13 & 66.86 & 5,474 & 2.017\\
mit         & 26.57 & 73.24 & 6,379 & 2.763\\
gatech      & 26.03 & 73.94 & 3,907 & 2.841\\
ufl         & 24.76 & 75.23 & 4,927 & 3.038\\
vt          & 23.07 & 76.92 & 4,061 & 3.334\\
lsu         & 15.06 & 84.93 & 2,974 & 5.638\\
nsu         & 13.82 & 86.00 & 1,208 & 6.233\\
odu         & 9.727 & 90.27 & 1,727 & 9.279\\
tcc         & 5.429 & 94.57 & 884 & 17.41\\
  \end{tabular}
  \caption{Dereferencing the TimeMaps from 13 academic institutions' Web sites from Internet Archive produces the above distribution of status codes for each site.}
  \label{tab:edu}
\end{table}

\begin{table}
\centering
\def\arraystretch{0}
\begin{tabular}{  c |  c  c  c  c }
\multirow{2}{*}{year} & \multicolumn{4}{c}{scheme$_{orig}\rightarrow $scheme$_{dest}$} \\ \cline{2-5}
& http$\rightarrow$http & \ooorange{http$\rightarrow$https} & https$\rightarrow$http & https$\rightarrow$https \\ \hline
 2010 & \includegraphics[width=0.16\linewidth]{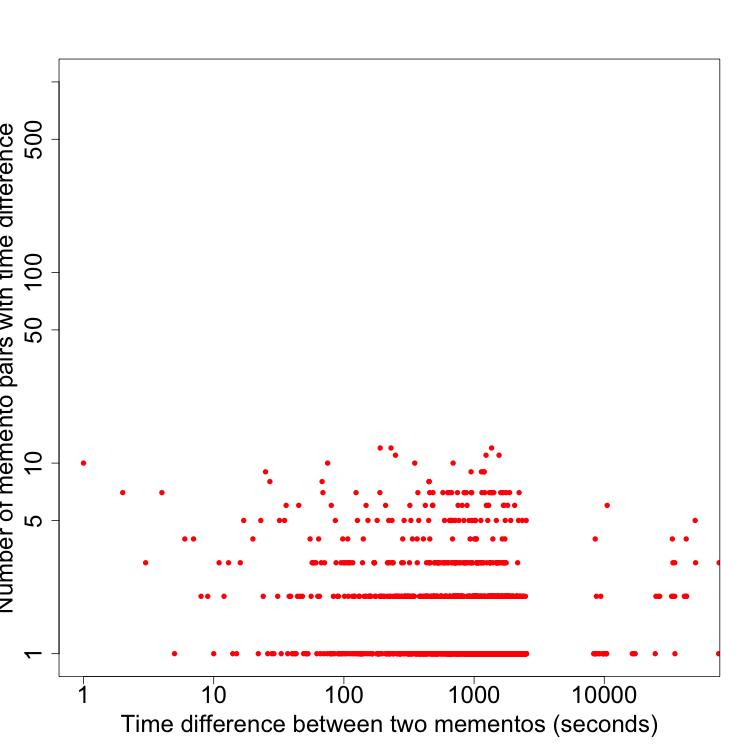} &   &  &  \\
 2011 & \includegraphics[width=0.16\linewidth]{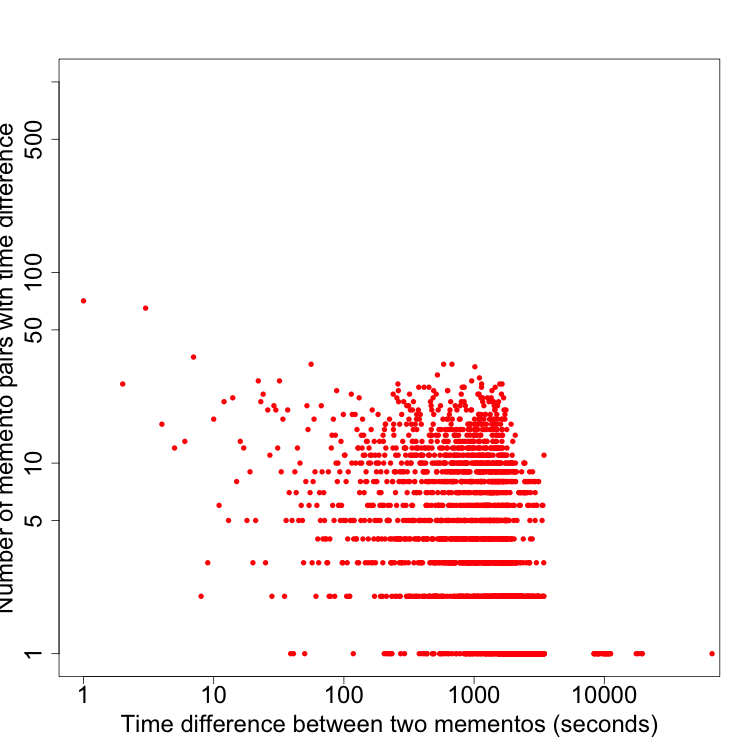} &   & \includegraphics[width=0.16\linewidth]{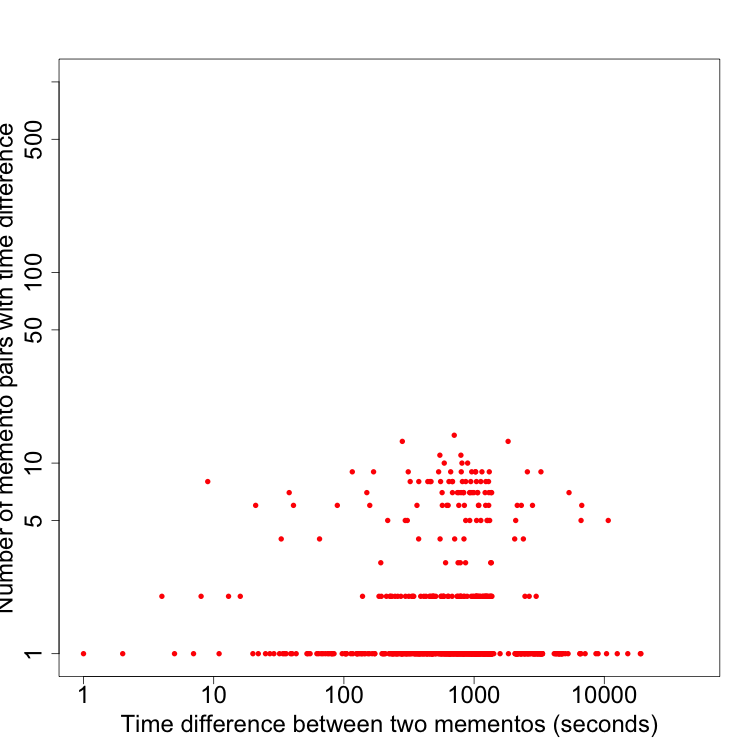} &  \\
 2012 & \includegraphics[width=0.16\linewidth]{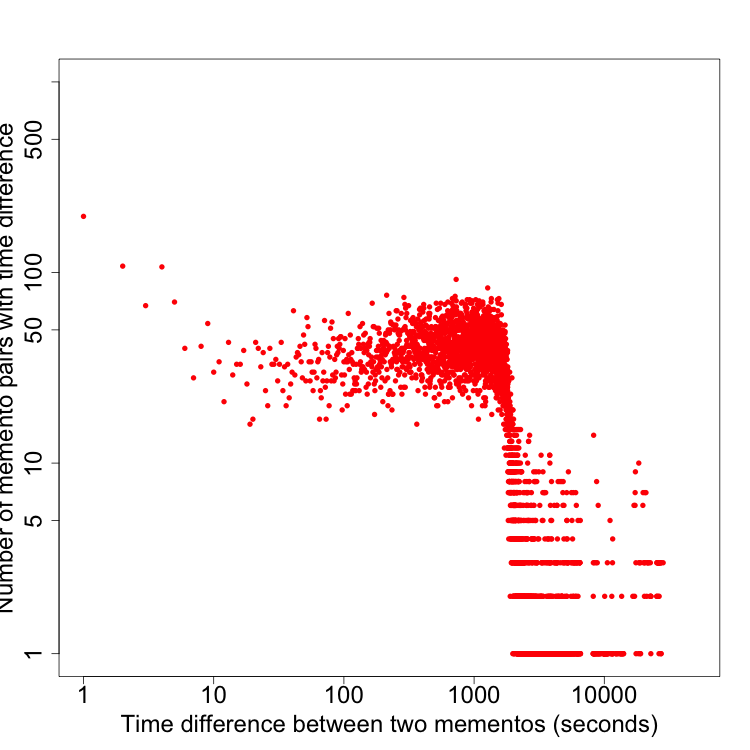} &  \includegraphics[width=0.16\linewidth]{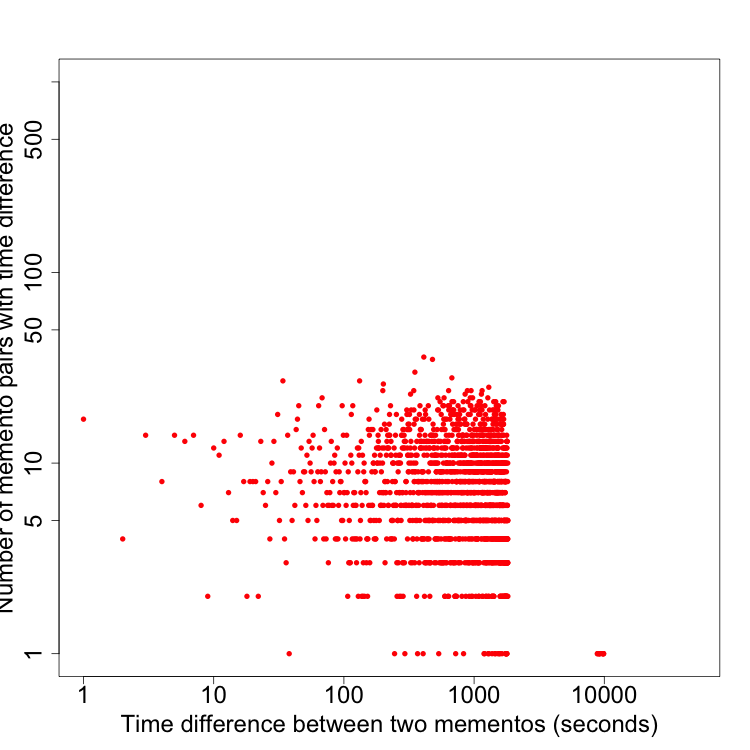}  & \includegraphics[width=0.16\linewidth]{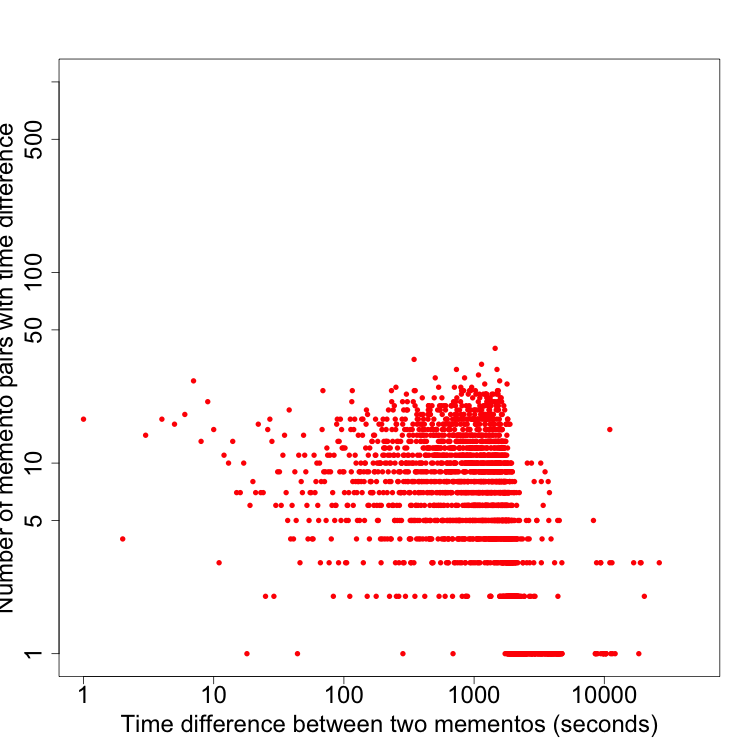}  & \includegraphics[width=0.16\linewidth]{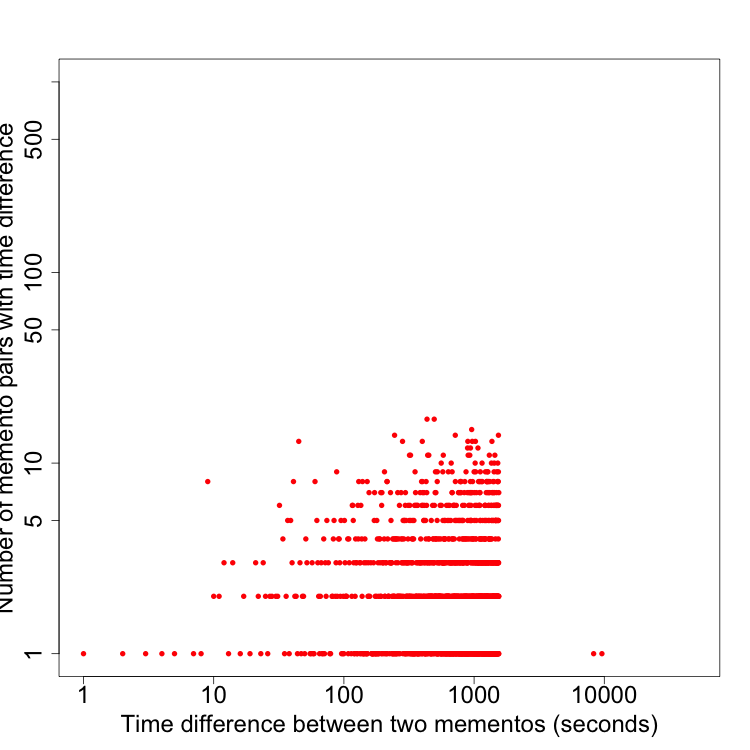} \\
 2013 & \includegraphics[width=0.16\linewidth]{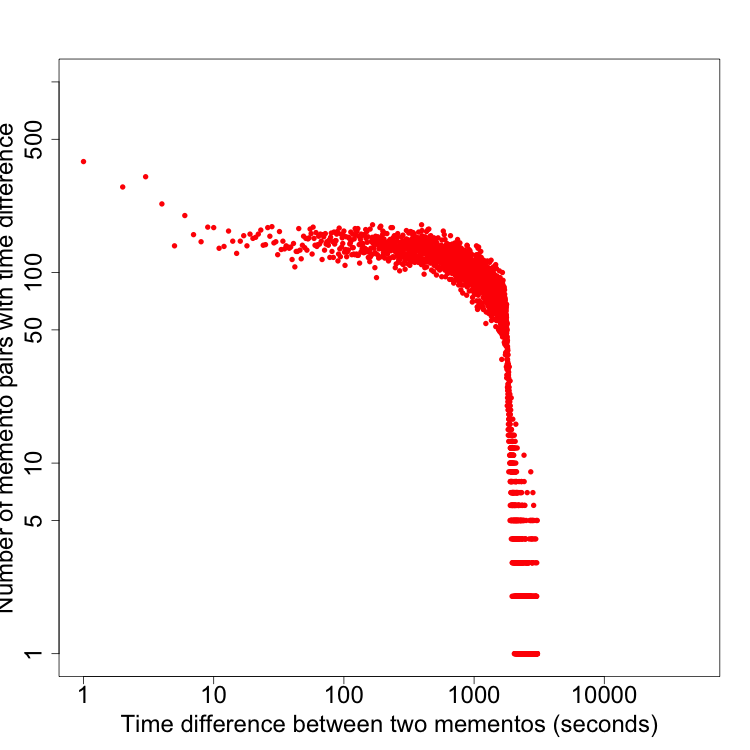} &  \includegraphics[width=0.16\linewidth]{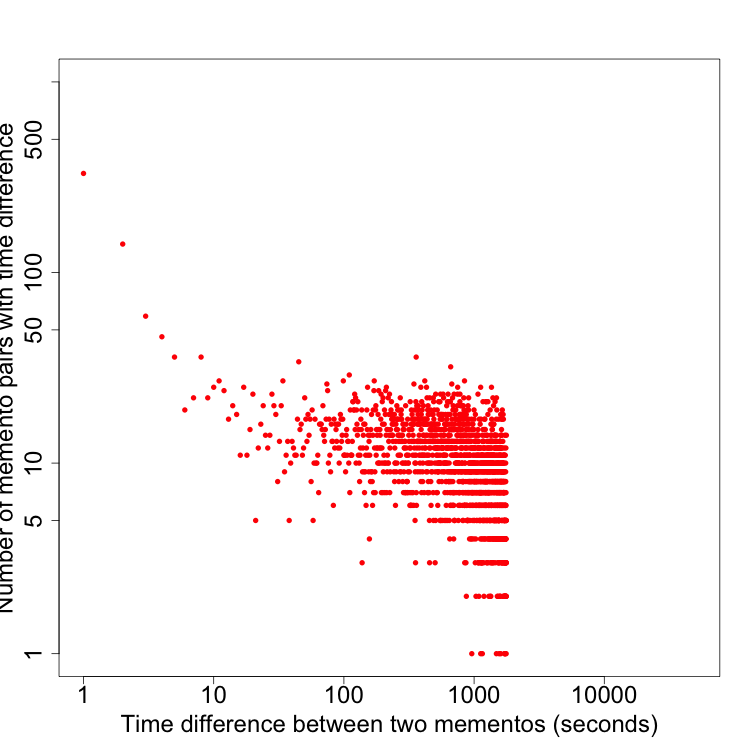}  & \includegraphics[width=0.16\linewidth]{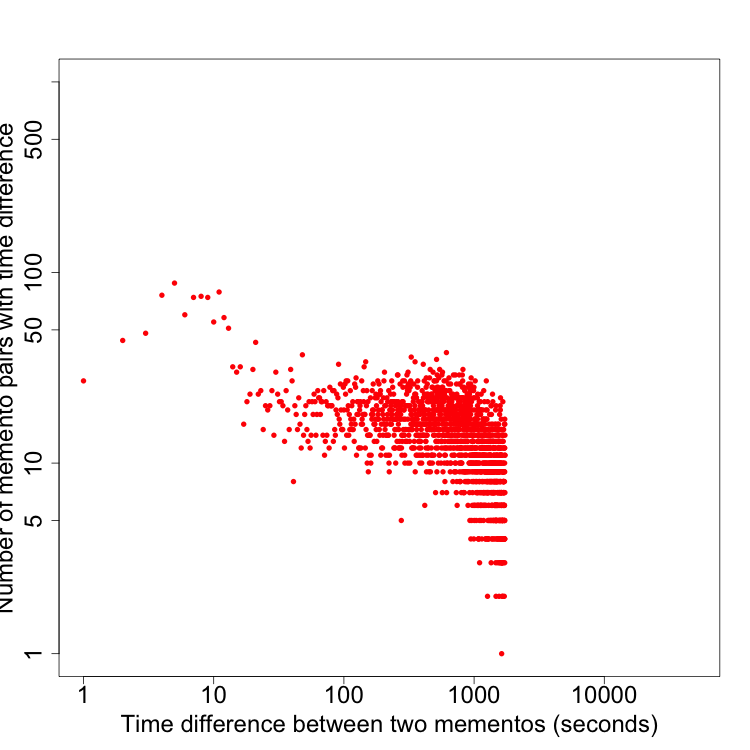}  & \includegraphics[width=0.16\linewidth]{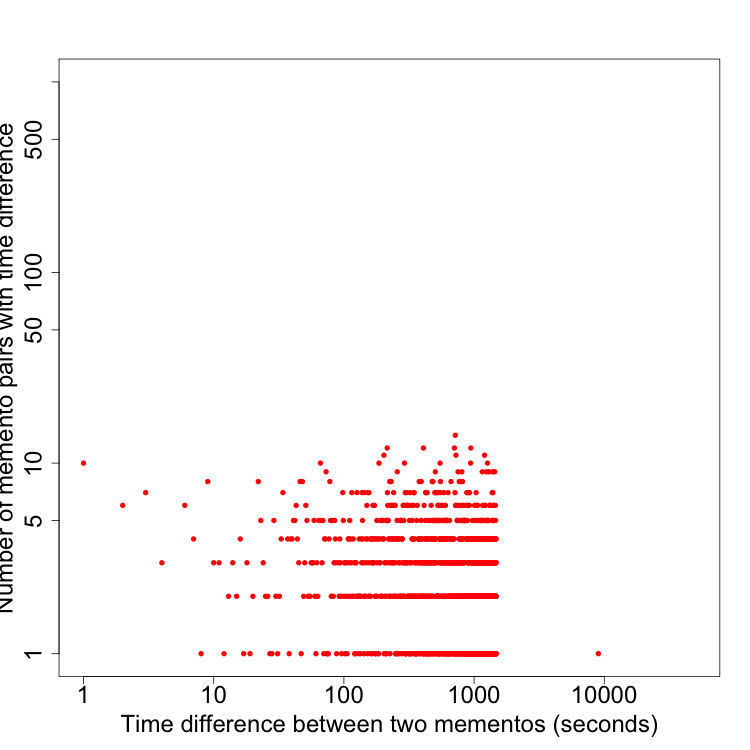} \\
 2014 & \includegraphics[width=0.16\linewidth]{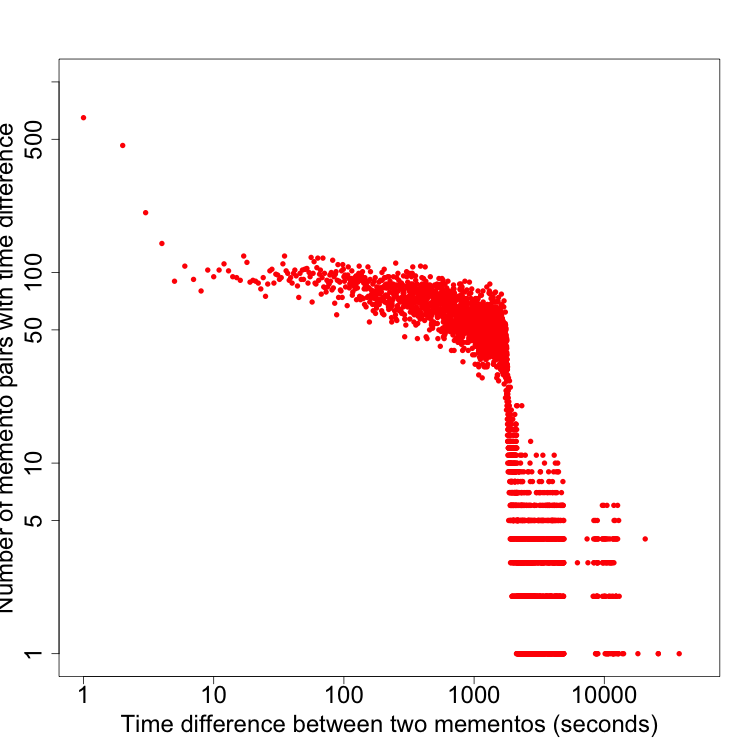} &  \includegraphics[width=0.16\linewidth]{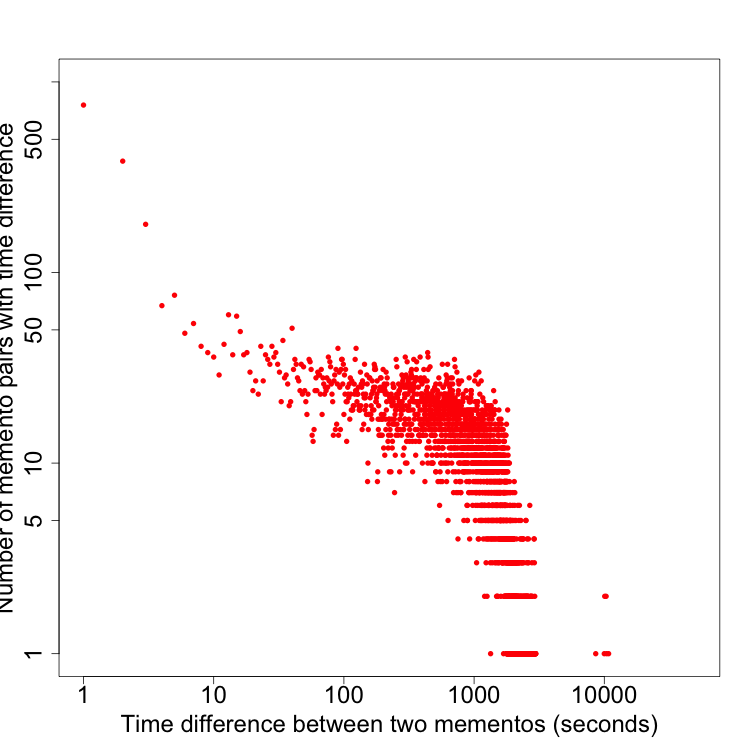}  & \includegraphics[width=0.16\linewidth]{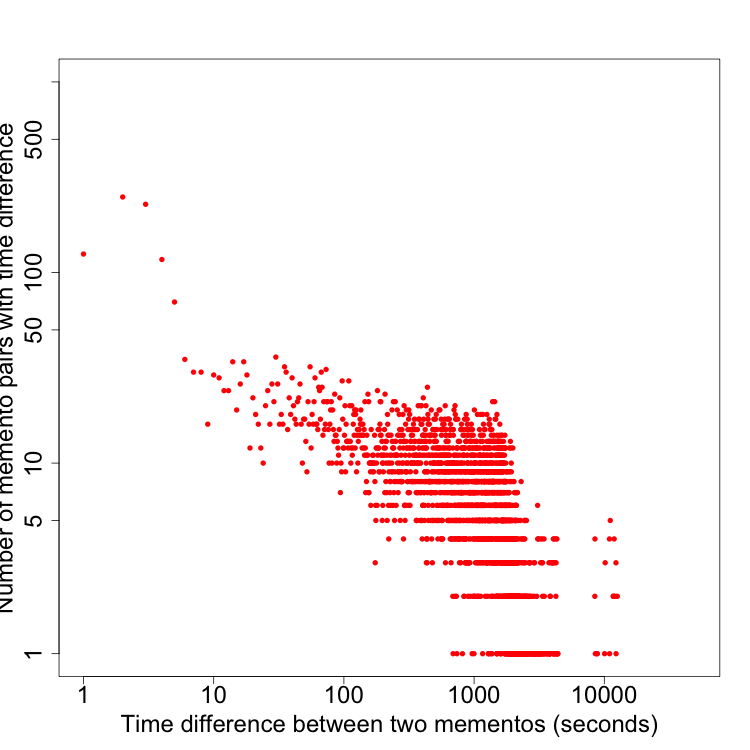}  & \includegraphics[width=0.16\linewidth]{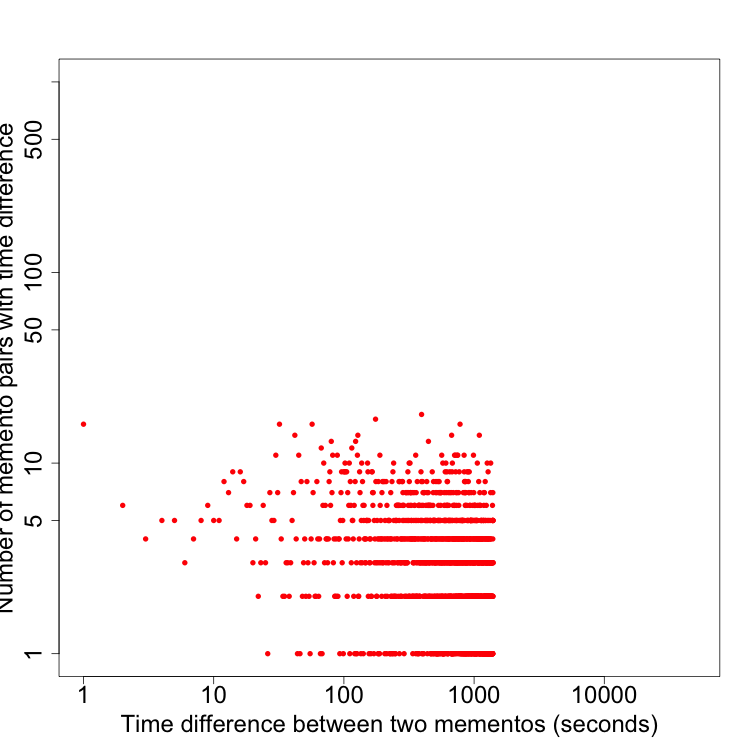} \\
 2015 & \includegraphics[width=0.16\linewidth]{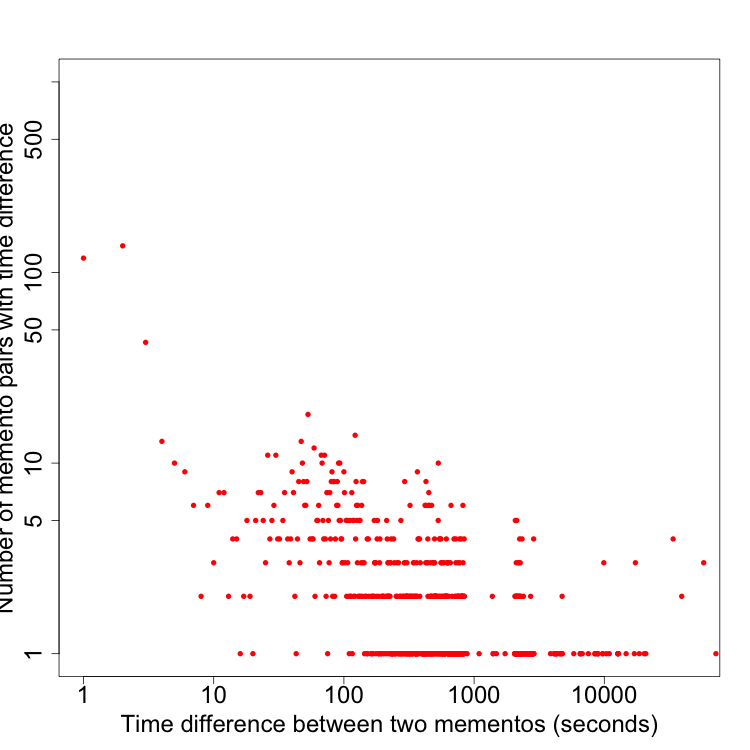} &
    &
 \includegraphics[width=0.16\linewidth]{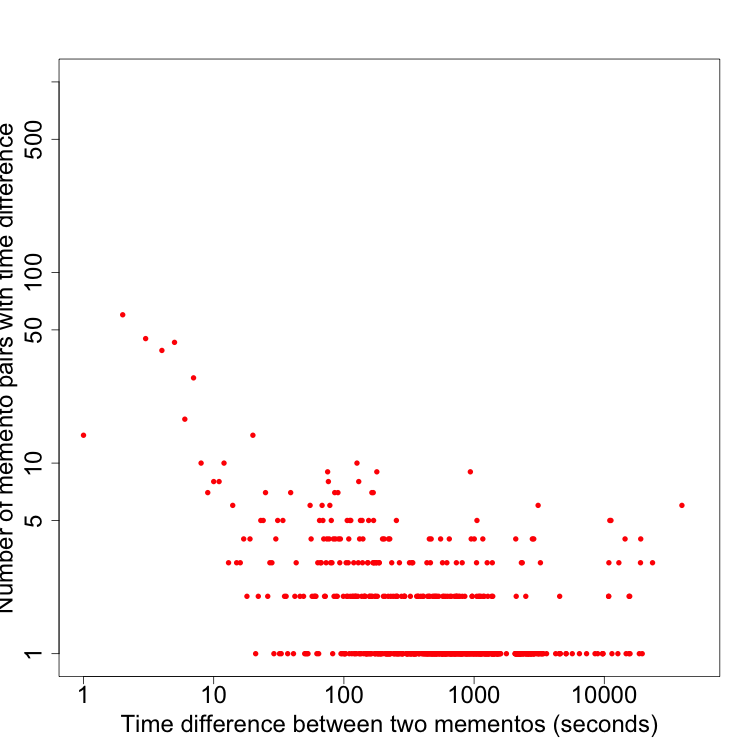}   &  \\
 2016 & \includegraphics[width=0.16\linewidth]{schemeForwardPlots/{archive.org_HTTPtoHTTP_2013_small}.png} &  \includegraphics[width=0.16\linewidth]{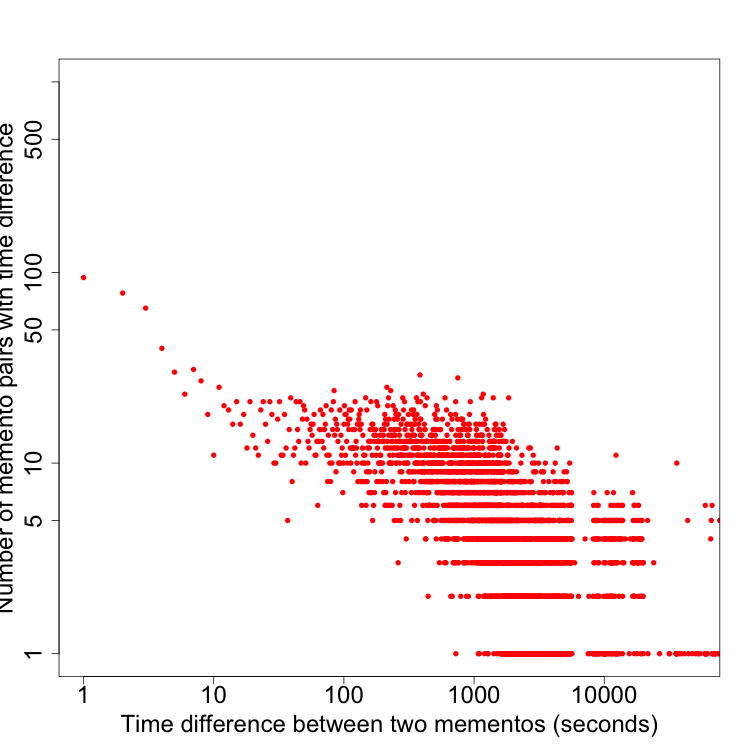}  & \includegraphics[width=0.16\linewidth]{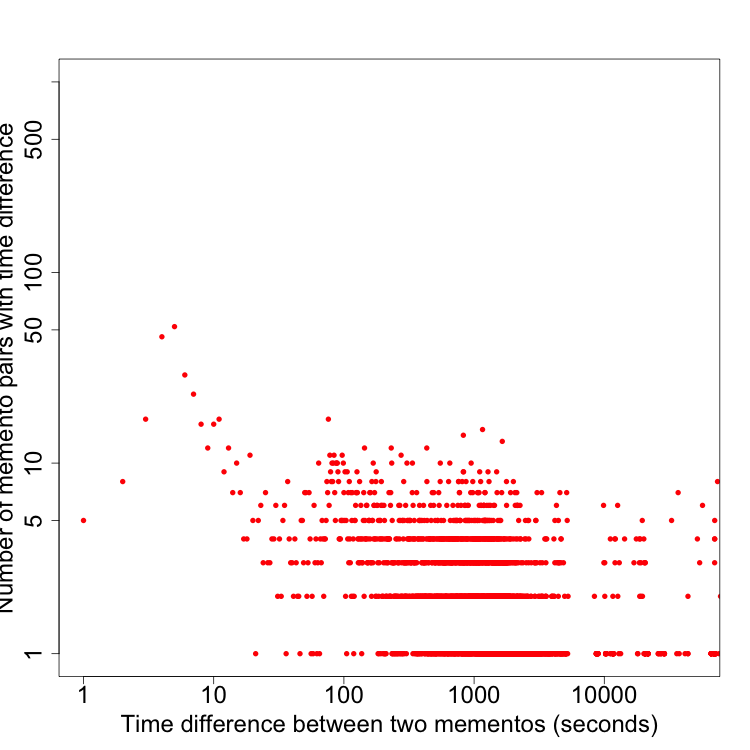}  & \includegraphics[width=0.16\linewidth]{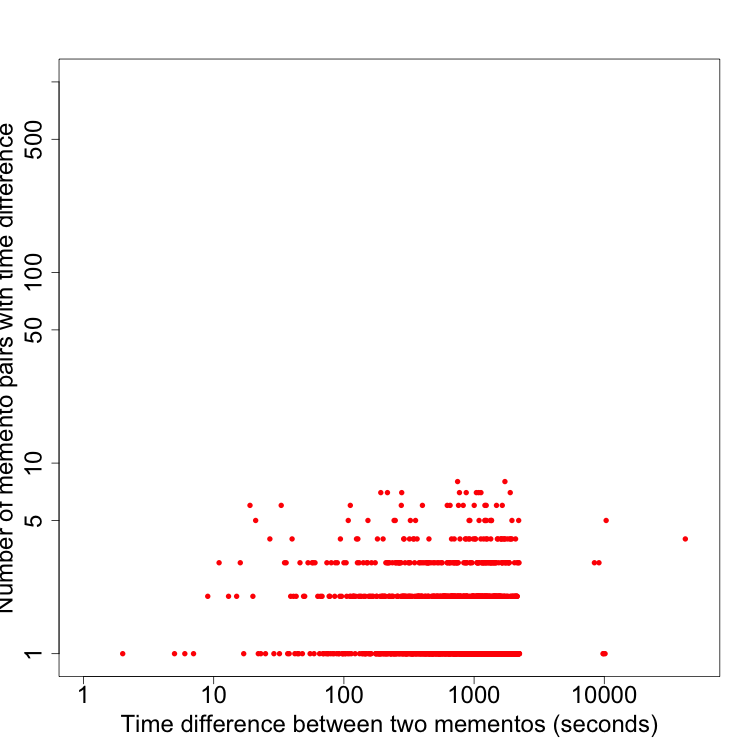} \\
\end{tabular}
\caption{Count of time delta instances for google.com from IA TimeMaps with 3XX redirects. Highlighted columns correspond to Figures~\ref{tab:datatatata} and \ref{fig:interScheme}.The plots in this Figure are available with more detail in the Appendix.
}
\label{tab:allThePlots}
\end{table}

\section{Conclusions}
In this work we identified the problem of attempting to count the number of mementos in a TimeMap based solely on the contents of the TimeMap. We progressively built a method for counting the number of archived captures of Web pages that contain content when dereferenced from a TimeMap. Through observing google.com, a \URIR with a contemporarily large apparent number of mementos, we dereferenced all URI-Ms in an aggregated TimeMap for the URI-R to show that a 84.9\% of the URI-Ms are redirects to other \URIMs in the TimeMap. 

We establish the nomenclature of M$_{RC}$, a means of communicating the number of URI-Ms in a TimeMap that contain a capture with an entity body, as compared to the more naive M$_{TM}$ as calculated using solely the contents of the TimeMap for a \URIRns. We analyzed the TimeMaps for the \URIRs of seven other contemporary large web sites and the TimeMaps from 13 academic institutions. We introduced the $DI$ metric to evaluate the ratio of non-redirecting \URIMs in a TimeMap to the ratio of redirecting \URIMs when all \URIMs in a TimeMap are dereferenced. Five of the eight large Web sites' \URIRs and two of the thirteen academic institutions'  \URIRs contained more redirecting than non-redirecting mementos when dereferenced ($DI < 1.0)$.

From the \URIMs for google.com that redirected, we split the results between those that changed schemes and those that maintained the same URI-R scheme after the redirect. We split the results on an annual basis to show the effect that the introduction of HTTPS has had on \URIR canonicalization over time. We found that despite an anomalous set of captures in the year 2015, the number of redirects per year on the live Web from HTTP \URIRs to HTTPS \URIRs as preserved by the archive has superseded the number of redirects of HTTPS \URIRs to HTTP \URIRsns. Though the quantity of holdings by Internet Archive for redirects from HTTPS to HTTP is not yet larger than the total of other permutations of HTTP(S) to HTTP(S) redirects (Table~\ref{tab:datatatata}), the rapid growth of redirects to the secure scheme for captures confirms and quantifies the increased adoption of HTTPS on the live Web.

\bibliographystyle{ACM-Reference-Format}
\bibliography{arXiv-countingMementos} 


\begin{thebibliography}{00}


\ifx \showCODEN    \undefined \def \showCODEN     #1{\unskip}     \fi
\ifx \showDOI      \undefined \def \showDOI       #1{{\tt DOI:}\penalty0{#1}\ }
  \fi
\ifx \showISBNx    \undefined \def \showISBNx     #1{\unskip}     \fi
\ifx \showISBNxiii \undefined \def \showISBNxiii  #1{\unskip}     \fi
\ifx \showISSN     \undefined \def \showISSN      #1{\unskip}     \fi
\ifx \showLCCN     \undefined \def \showLCCN      #1{\unskip}     \fi
\ifx \shownote     \undefined \def \shownote      #1{#1}          \fi
\ifx \showarticletitle \undefined \def \showarticletitle #1{#1}   \fi
\ifx \showURL      \undefined \def \showURL       #1{#1}          \fi
\providecommand\bibfield[2]{#2}
\providecommand\bibinfo[2]{#2}
\providecommand\natexlab[1]{#1}
\providecommand\showeprint[2][]{arXiv:#2}

\bibitem[\protect\citeauthoryear{Alam and Nelson}{Alam and Nelson}{2016}]%
        {memgator}
\bibfield{author}{\bibinfo{person}{Sawood Alam} {and}
  \bibinfo{person}{Michael~L. Nelson}.} \bibinfo{year}{2016}\natexlab{}.
\newblock \showarticletitle{{MemGator - A Portable Concurrent Memento
  Aggregator}}. In \bibinfo{booktitle}{{\em Proceedings of the 16th ACM/IEEE-CS
  on Joint Conference on Digital Libraries (JCDL)}}. \bibinfo{pages}{243--244}.
\newblock
\showDOI{%
\url{http://dx.doi.org/10.1145/2910896.2925452}}


\bibitem[\protect\citeauthoryear{Alam, Nelson, {Van de Sompel}, Balakireva,
  Shankar, and Rosenthal}{Alam et~al\mbox{.}}{2015}]%
        {salam-cdxj}
\bibfield{author}{\bibinfo{person}{Sawood Alam}, \bibinfo{person}{Michael~L.
  Nelson}, \bibinfo{person}{Herbert {Van de Sompel}}, \bibinfo{person}{Lyudmila
  Balakireva}, \bibinfo{person}{Harihar Shankar}, {and} \bibinfo{person}{David
  S.~H. Rosenthal}.} \bibinfo{year}{2015}\natexlab{}.
\newblock \showarticletitle{{Web Archive Profiling Through CDX Summarization}}.
  In \bibinfo{booktitle}{{\em Proceedings of Theory and Practice of Digital
  Libraries (TPDL)}}. \bibinfo{pages}{3--14}.
\newblock
\showDOI{%
\url{http://dx.doi.org/10.1007/s00799-016-0184-4}}


\bibitem[\protect\citeauthoryear{{AlSum}, Sanderson, {Van de Sompel}, and
  Nelson}{{AlSum} et~al\mbox{.}}{2013}]%
        {alsum-redirects}
\bibfield{author}{\bibinfo{person}{Ahmed {AlSum}}, \bibinfo{person}{Robert
  Sanderson}, \bibinfo{person}{Herbert {Van de Sompel}}, {and}
  \bibinfo{person}{Michael~L. Nelson}.} \bibinfo{year}{2013}\natexlab{}.
\newblock \showarticletitle{Archival {HTTP} Redirection Retrieval Policies}. In
  \bibinfo{booktitle}{{\em Proceedings of the Third Temporal Web Analytics
  Workshop}}.
\newblock
\showDOI{%
\url{http://dx.doi.org/10.1145/2487788.2488117}}


\bibitem[\protect\citeauthoryear{Bar-Yossef, Broder, Kumar, and
  Tompkins}{Bar-Yossef et~al\mbox{.}}{2004}]%
        {bar-yossef}
\bibfield{author}{\bibinfo{person}{Ziv Bar-Yossef}, \bibinfo{person}{Andrei~Z.
  Broder}, \bibinfo{person}{Ravi Kumar}, {and} \bibinfo{person}{Andrew
  Tompkins}.} \bibinfo{year}{2004}\natexlab{}.
\newblock \showarticletitle{{Sic Transit Gloria Telae: Towards an Understanding
  of the Web's Decay}}. In \bibinfo{booktitle}{{\em {Proceedings of the 13th
  International Conference on World Wide Web (WWW)}}}.
  \bibinfo{pages}{328--337}.
\newblock
\showDOI{%
\url{http://dx.doi.org/10.1145/988672.988716}}


\bibitem[\protect\citeauthoryear{Fielding and Reschke}{Fielding and
  Reschke}{2014a}]%
        {rfc7230}
\bibfield{author}{\bibinfo{person}{R. Fielding} {and} \bibinfo{person}{J.
  Reschke}.} \bibinfo{year}{2014}\natexlab{a}.
\newblock \bibinfo{title}{{Hypertext Transfer Protocol (HTTP/1.1): Message
  Syntax and Routing, Internet RFC-7230}}.
\newblock   (\bibinfo{year}{2014}).
\newblock


\bibitem[\protect\citeauthoryear{Fielding and Reschke}{Fielding and
  Reschke}{2014b}]%
        {rfc7231}
\bibfield{author}{\bibinfo{person}{R. Fielding} {and} \bibinfo{person}{J.
  Reschke}.} \bibinfo{year}{2014}\natexlab{b}.
\newblock \bibinfo{title}{{Hypertext Transfer Protocol (HTTP/1.1): Semantics
  and Content, Internet RFC-7231}}.
\newblock   (\bibinfo{year}{2014}).
\newblock


\bibitem[\protect\citeauthoryear{Jacobs and Walsh}{Jacobs and Walsh}{2004}]%
        {webArchitecture-opacity}
\bibfield{author}{\bibinfo{person}{Ian Jacobs} {and} \bibinfo{person}{Norman
  Walsh}.} \bibinfo{year}{2004}\natexlab{}.
\newblock \bibinfo{title}{{Web Architecture : URI Opacity}}.
\newblock \bibinfo{howpublished}{https://www.w3.org/TR/webarch/\#uri-opacity}.
   (\bibinfo{year}{2004}).
\newblock
\showURL{%
\url{https://www.w3.org/TR/webarch/\#uri-opacity}}


\bibitem[\protect\citeauthoryear{Jordan, Kelly, Brunelle, Vobrak, Weigle, and
  Nelson}{Jordan et~al\mbox{.}}{2015}]%
        {jordan-jcdl2015}
\bibfield{author}{\bibinfo{person}{Wesley Jordan}, \bibinfo{person}{Mat Kelly},
  \bibinfo{person}{Justin~F. Brunelle}, \bibinfo{person}{Laura Vobrak},
  \bibinfo{person}{Michele~C. Weigle}, {and} \bibinfo{person}{Michael~L.
  Nelson}.} \bibinfo{year}{2015}\natexlab{}.
\newblock \showarticletitle{{ Mobile Mink: Merging Mobile and Desktop Archived
  Webs }}. In \bibinfo{booktitle}{{\em Proceedings of the ACM/IEEE Joint
  Conference on Digital Libraries (JCDL)}}. \bibinfo{pages}{243--244}.
\newblock
\showDOI{%
\url{http://dx.doi.org/10.1145/2756406.2756956}}


\bibitem[\protect\citeauthoryear{Kelly, Nelson, and Weigle}{Kelly
  et~al\mbox{.}}{2014}]%
        {kelly-dl2014-mink}
\bibfield{author}{\bibinfo{person}{Mat Kelly}, \bibinfo{person}{Michael~L.
  Nelson}, {and} \bibinfo{person}{Michele~C. Weigle}.}
  \bibinfo{year}{2014}\natexlab{}.
\newblock \showarticletitle{{Mink: Integrating the Live and Archived Web
  Viewing Experience Using Web Browsers and Memento}}. In
  \bibinfo{booktitle}{{\em Proceedings of the IEEE/ACM Joint Conference on
  Digital Libraries (JCDL)}}. \bibinfo{pages}{469--470}.
\newblock
\showDOI{%
\url{http://dx.doi.org/10.1109/JCDL.2014.6970229}}


\bibitem[\protect\citeauthoryear{Meneses, Furuta, and Shipman}{Meneses
  et~al\mbox{.}}{2012}]%
        {soft404s}
\bibfield{author}{\bibinfo{person}{Luis Meneses}, \bibinfo{person}{Richard
  Furuta}, {and} \bibinfo{person}{Frank Shipman}.}
  \bibinfo{year}{2012}\natexlab{}.
\newblock \showarticletitle{{Identifying ``Soft 404'' Error Pages: Analyzing
  the Lexical Signatures of Documents in Distributed Collections}}. In
  \bibinfo{booktitle}{{\em Proceedings of the International Conference on
  Theory and Practice of Digital Libraries (TPDL)}}. \bibinfo{pages}{197--208}.
\newblock
\showDOI{%
\url{http://dx.doi.org/10.1007/978-3-642-33290-6_22}}


\bibitem[\protect\citeauthoryear{Nelson}{Nelson}{2013}]%
        {wsdl-ia-memento}
\bibfield{author}{\bibinfo{person}{Michael~L. Nelson}.}
  \bibinfo{year}{2013}\natexlab{}.
\newblock \bibinfo{title}{{Wayback Machine Upgrades Memento Support}}.
\newblock
  \bibinfo{howpublished}{http://ws-dl.blogspot.com/2013/07/2013-07-15-wayback-machine-upgrades.html}.
    (\bibinfo{date}{September} \bibinfo{year}{2013}).
\newblock


\bibitem[\protect\citeauthoryear{Ohye and Kupke}{Ohye and Kupke}{2012}]%
        {rfc6596}
\bibfield{author}{\bibinfo{person}{M. Ohye} {and} \bibinfo{person}{J. Kupke}.}
  \bibinfo{year}{2012}\natexlab{}.
\newblock \bibinfo{title}{{The Canonical Link Relation, Internet RFC-6596}}.
\newblock   (\bibinfo{year}{2012}).
\newblock


\bibitem[\protect\citeauthoryear{Podjarny}{Podjarny}{2016}]%
        {httpsBreaking}
\bibfield{author}{\bibinfo{person}{Guy Podjarny}.}
  \bibinfo{year}{2016}\natexlab{}.
\newblock \bibinfo{title}{{HTTPS Adoption *doubled* this year | Snyk}}.
\newblock   (\bibinfo{date}{July} \bibinfo{year}{2016}).
\newblock
\showURL{%
\url{https://snyk.io/blog/https-breaking-through/}}
\newblock
\shownote{[Online; accessed 21-January-2017].}


\bibitem[\protect\citeauthoryear{Robinson and Timm}{Robinson and Timm}{2015}]%
        {freedompress}
\bibfield{author}{\bibinfo{person}{Garrett Robinson} {and}
  \bibinfo{person}{Trevor Timm}.} \bibinfo{year}{2015}\natexlab{}.
\newblock \bibinfo{title}{Introducing Secure The News, an automated tool
  tracking the adoption of HTTPS encryption across news websites}.
\newblock   (\bibinfo{date}{December} \bibinfo{year}{2015}).
\newblock
\showURL{%
\url{https://freedom.press/news/introducing-secure-news-automated-tool-tracking-adoption-https-encryption-across-news-websites/}}
\newblock
\shownote{[Online; accessed 21-January-2017].}


\bibitem[\protect\citeauthoryear{Rosenthal}{Rosenthal}{2013}]%
        {rosenthal-mementoAggregation}
\bibfield{author}{\bibinfo{person}{David S.~H. Rosenthal}.}
  \bibinfo{year}{2013}\natexlab{}.
\newblock \bibinfo{title}{{Re-thinking Memento Aggregation}}.
\newblock
  \bibinfo{howpublished}{\url{http://blog.dshr.org/2013/03/re-thinking-memento-aggregation.html}}.
    (\bibinfo{date}{March} \bibinfo{year}{2013}).
\newblock


\bibitem[\protect\citeauthoryear{Van~de Sompel, Nelson, and Sanderson}{Van~de
  Sompel et~al\mbox{.}}{2013}]%
        {rfc7089}
\bibfield{author}{\bibinfo{person}{Herbert Van~de Sompel},
  \bibinfo{person}{Michael Nelson}, {and} \bibinfo{person}{Robert Sanderson}.}
  \bibinfo{year}{2013}\natexlab{}.
\newblock \bibinfo{title}{{HTTP Framework for Time-Based Access to Resource
  States -- Memento}}.
\newblock \bibinfo{howpublished}{{IETF RFC 7089}}.
  (\bibinfo{date}{{December}} \bibinfo{year}{2013}).
\newblock


\end{thebibliography}

\clearpage
\appendix
\section{Detailed plots from Table 10} 
\label{appendix:largePlots}

\begin{figure}
\centering
\includegraphics[width=1.0\linewidth]{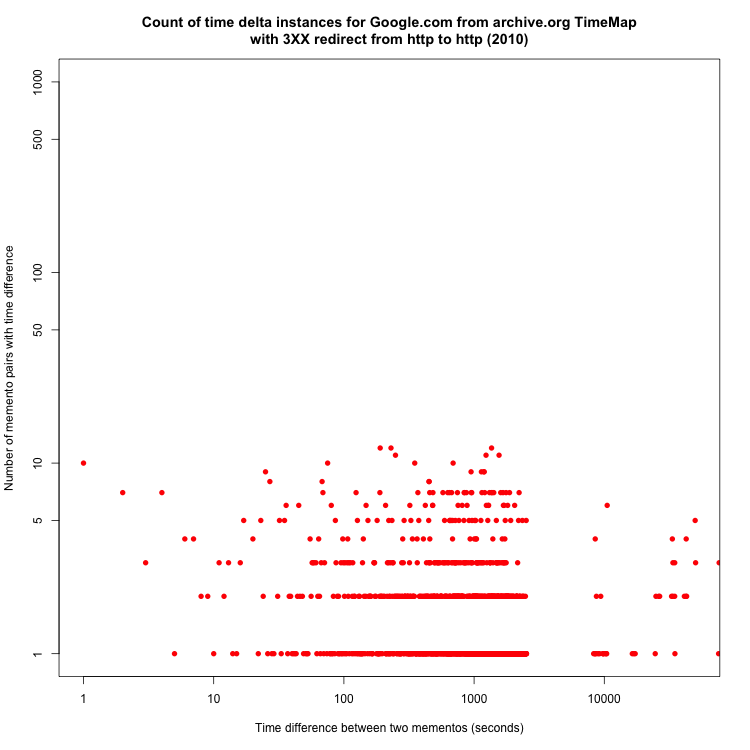}
\end{figure}

\begin{figure}
\centering
\includegraphics[width=1.0\linewidth]{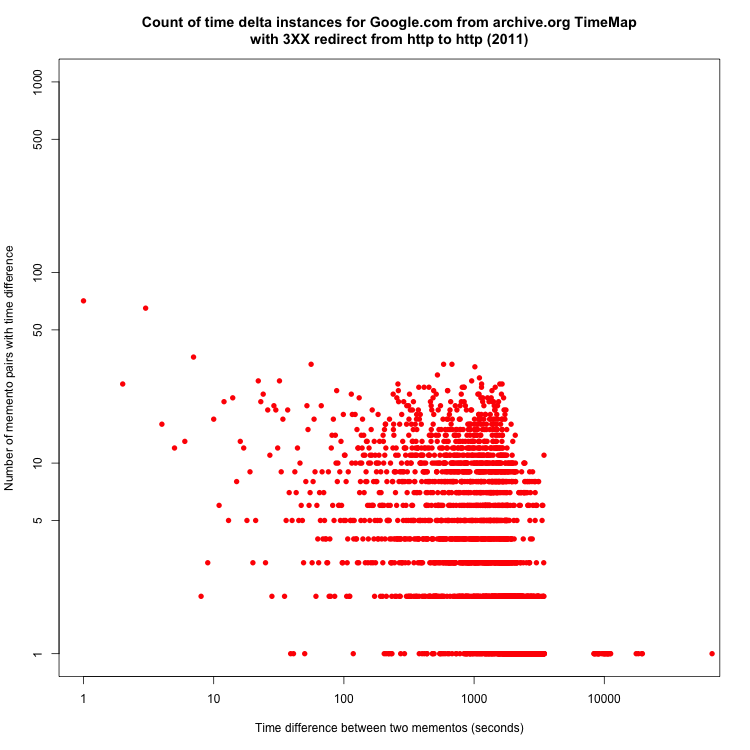} 
\end{figure}

\begin{figure}
\centering
\includegraphics[width=1.0\linewidth]{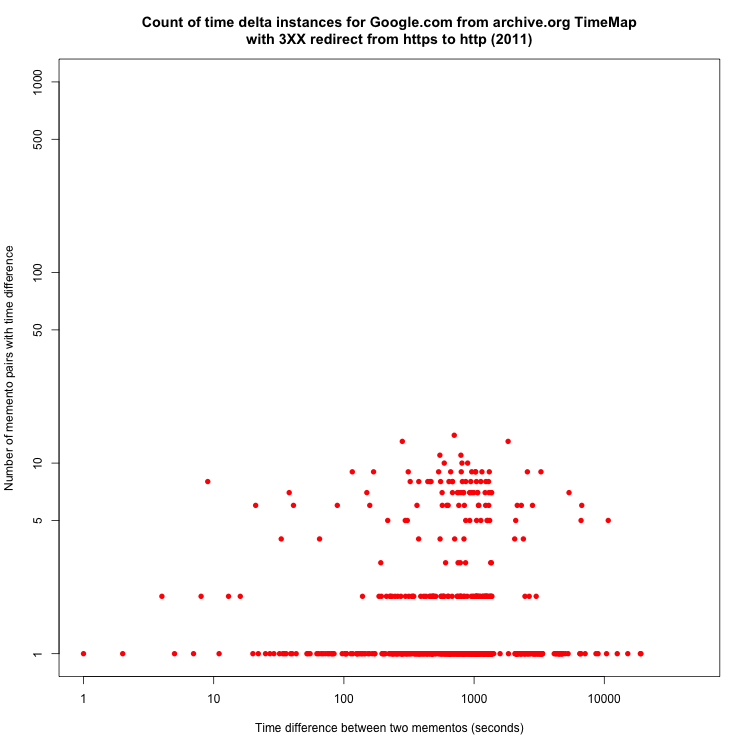} 
\end{figure}

\begin{figure}
\centering
\includegraphics[width=1.0\linewidth]{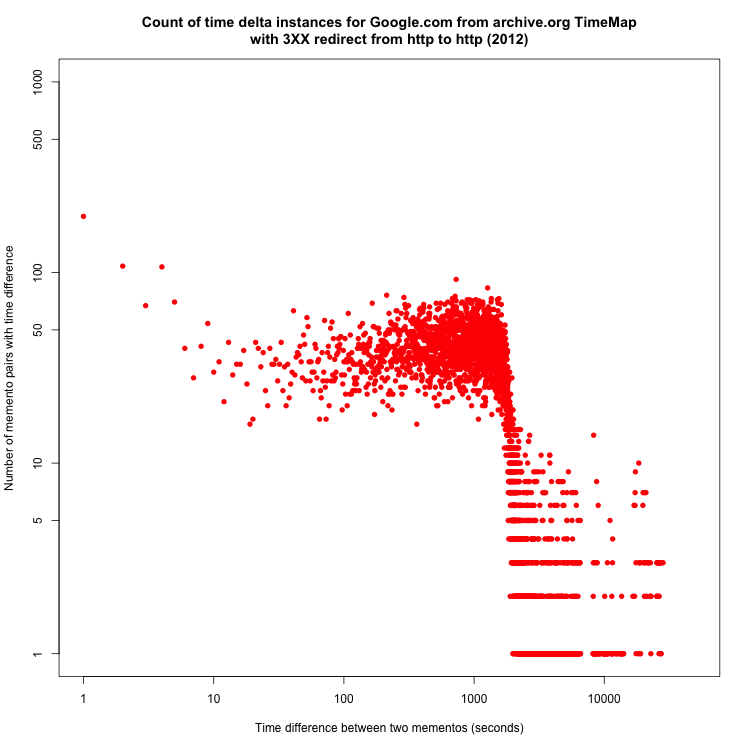}  
\end{figure}

\begin{figure}
\centering
\includegraphics[width=1.0\linewidth]{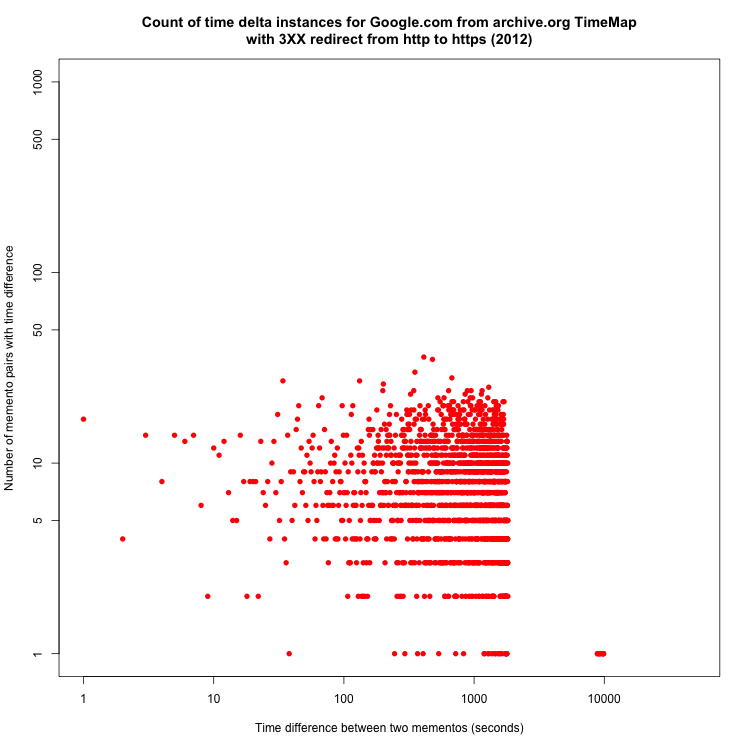}  
\end{figure}

\begin{figure}
\centering
\includegraphics[width=1.0\linewidth]{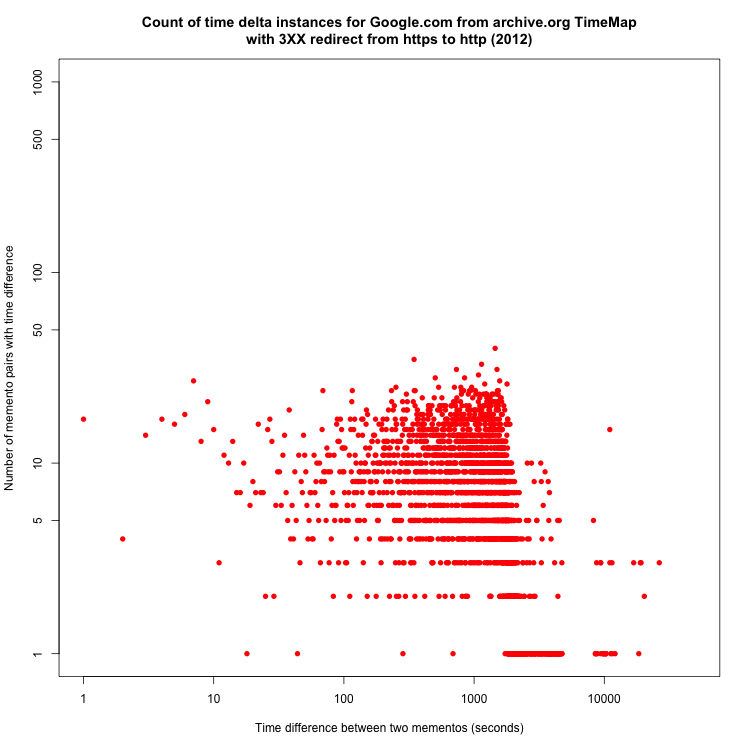}  
\end{figure}

\begin{figure}
\centering
\includegraphics[width=1.0\linewidth]{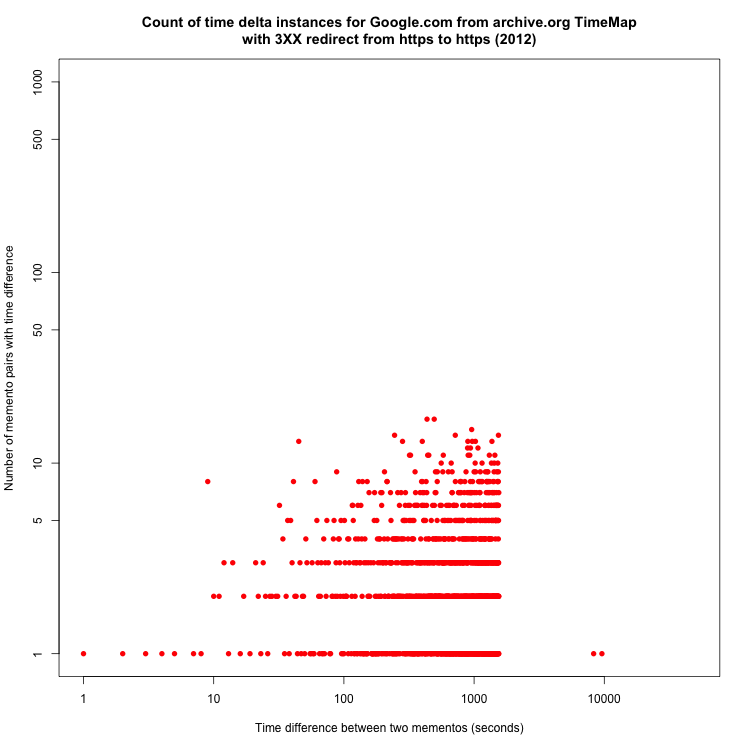} 
\end{figure}

\begin{figure}
\centering
\includegraphics[width=1.0\linewidth]{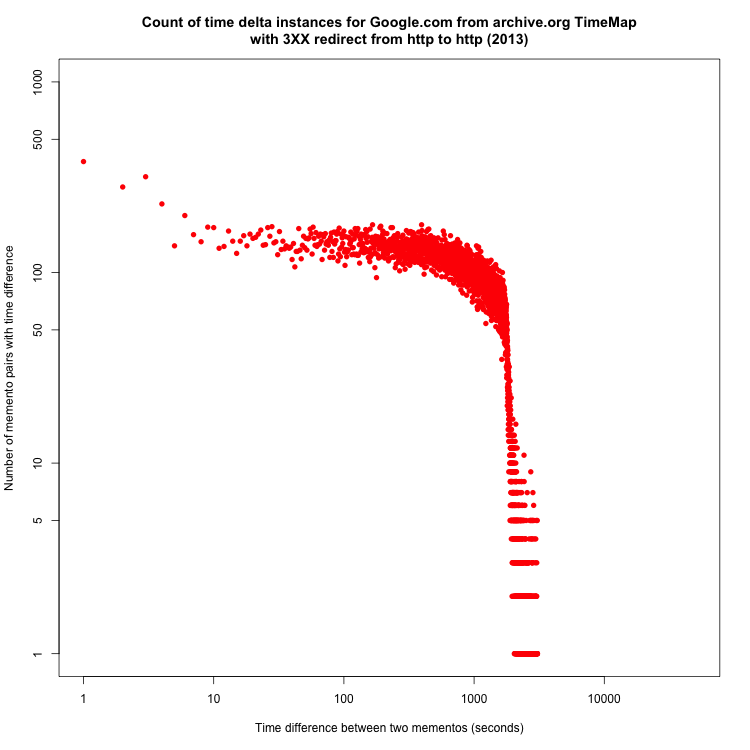} 
\end{figure}

\begin{figure}
\centering
\includegraphics[width=1.0\linewidth]{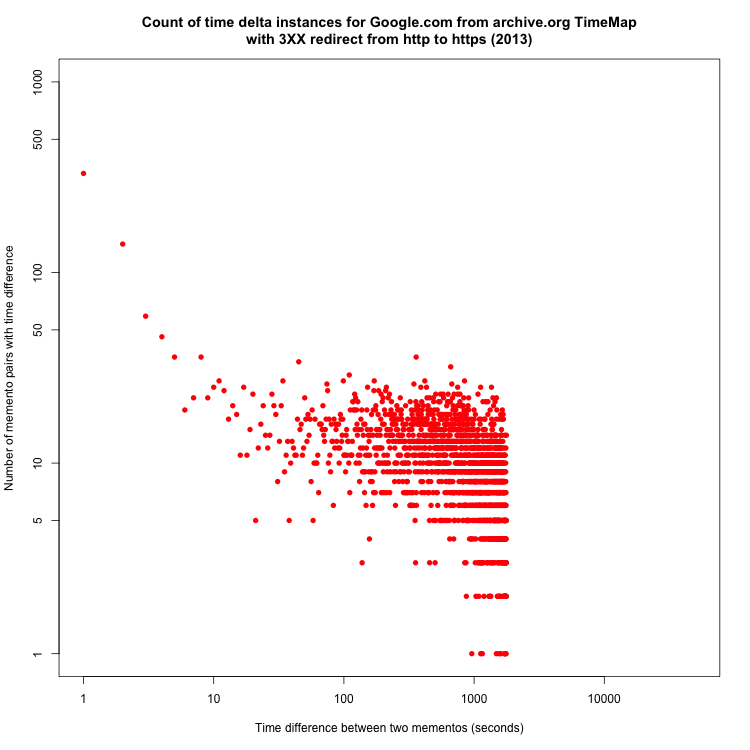}  
\end{figure}

\begin{figure}
\centering
\includegraphics[width=1.0\linewidth]{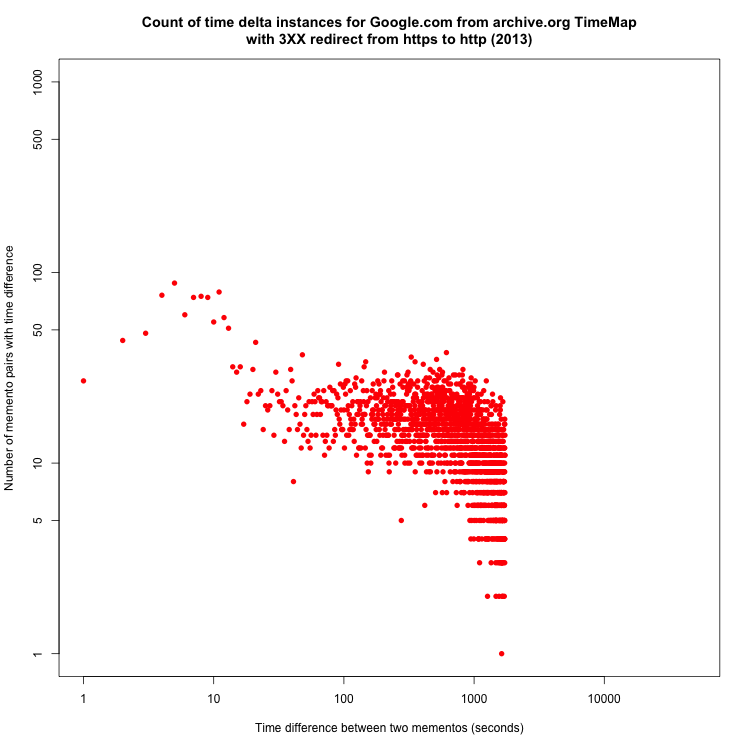}  
\end{figure}

\begin{figure}
\centering
\includegraphics[width=1.0\linewidth]{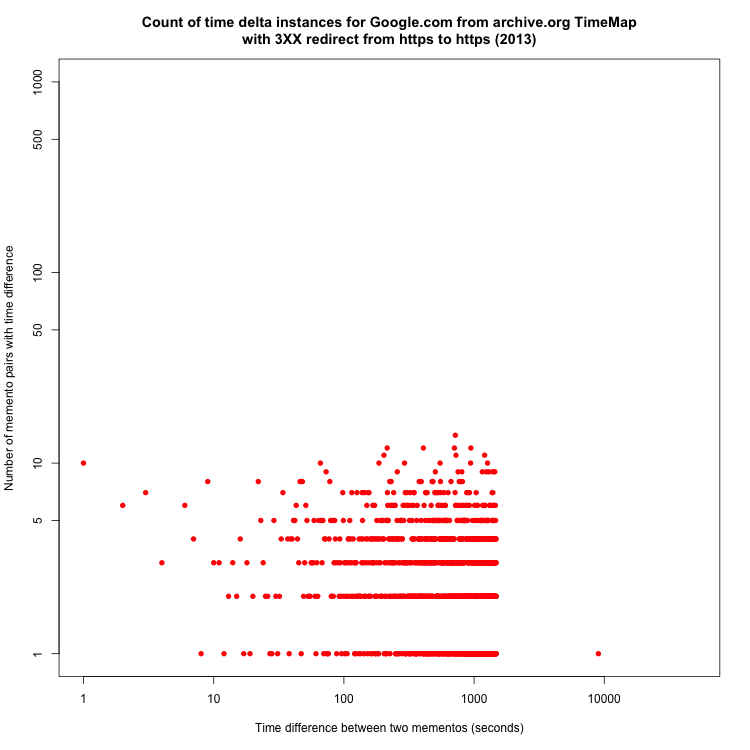} 
\end{figure}

\begin{figure}
\centering
\includegraphics[width=1.0\linewidth]{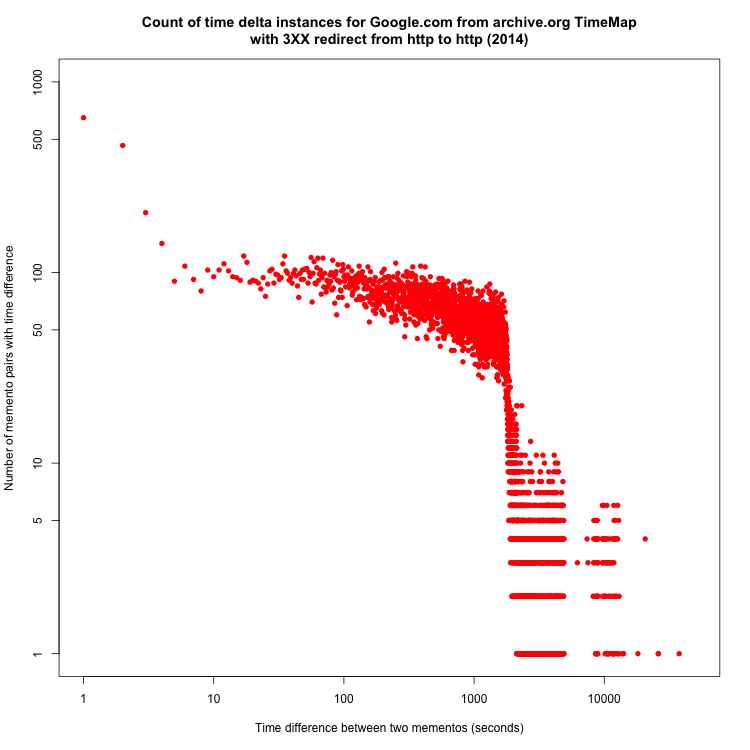}  
\end{figure}

\begin{figure}
\centering
\includegraphics[width=1.0\linewidth]{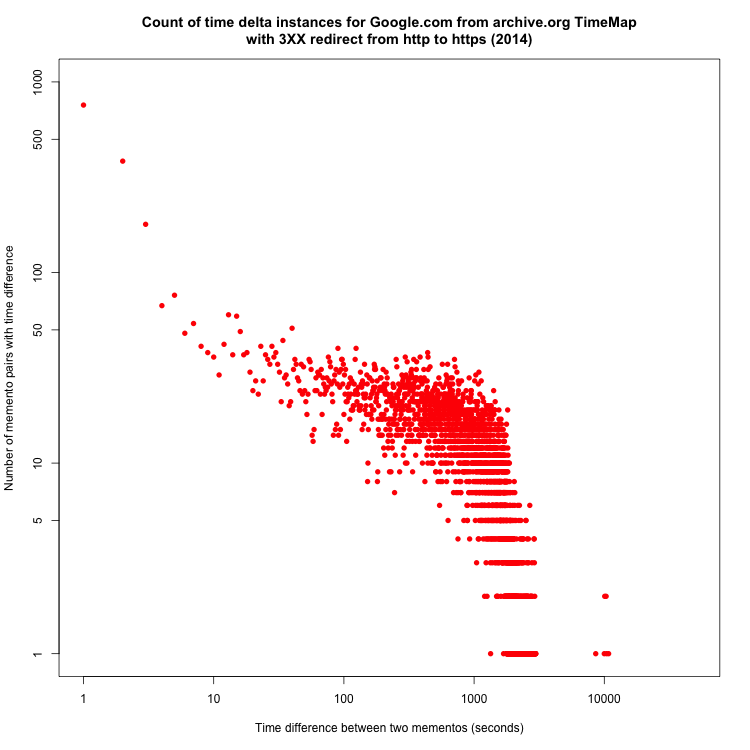}  
\end{figure}

\begin{figure}
\centering
\includegraphics[width=1.0\linewidth]{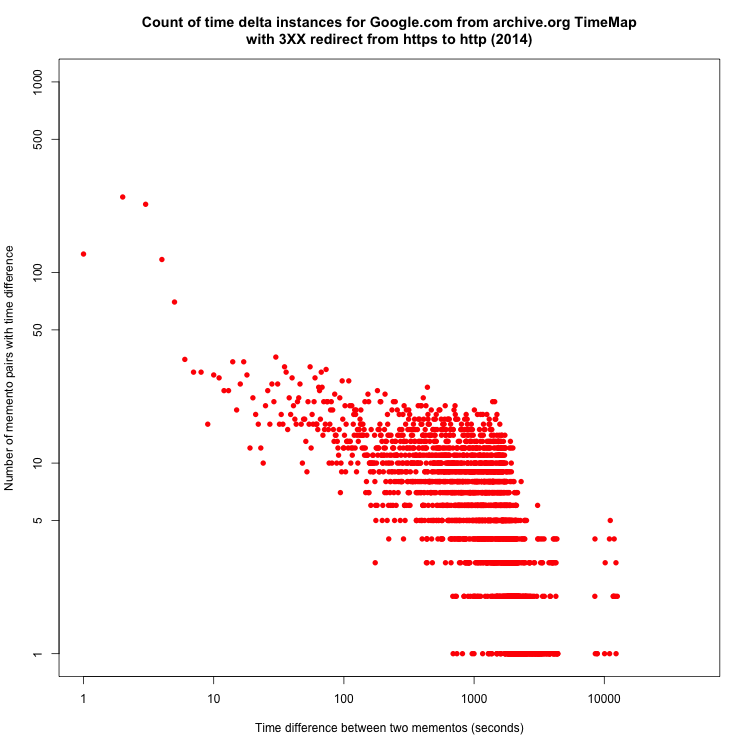} 
\end{figure}

\begin{figure}
\centering
\includegraphics[width=1.0\linewidth]{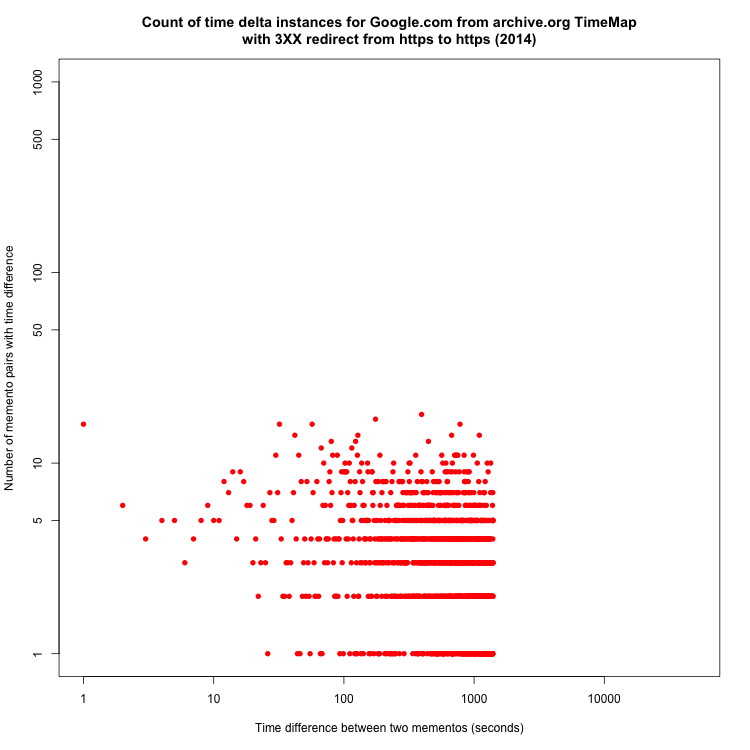} 
\end{figure}

\begin{figure}
\centering
\includegraphics[width=1.0\linewidth]{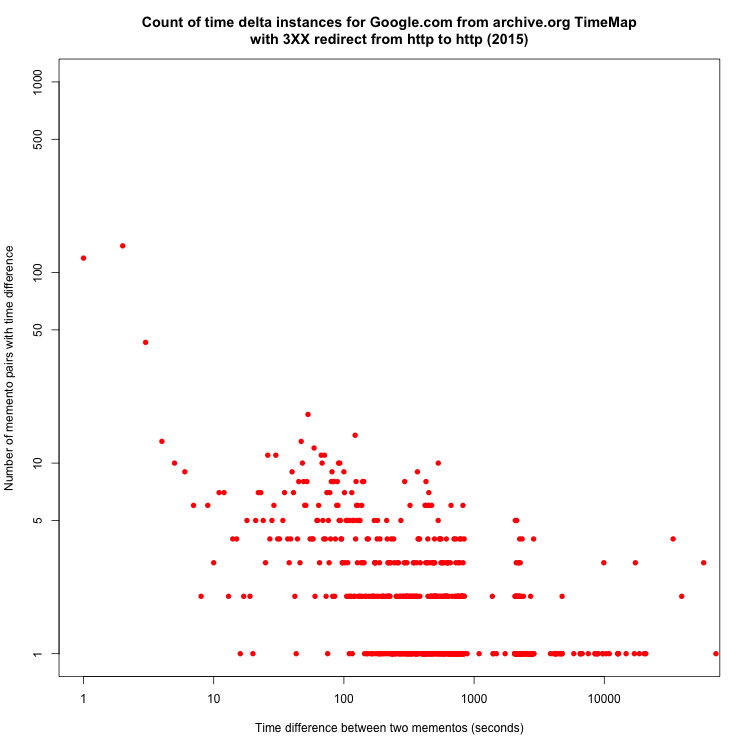}
\end{figure}

\begin{figure}
\centering
\includegraphics[width=1.0\linewidth]{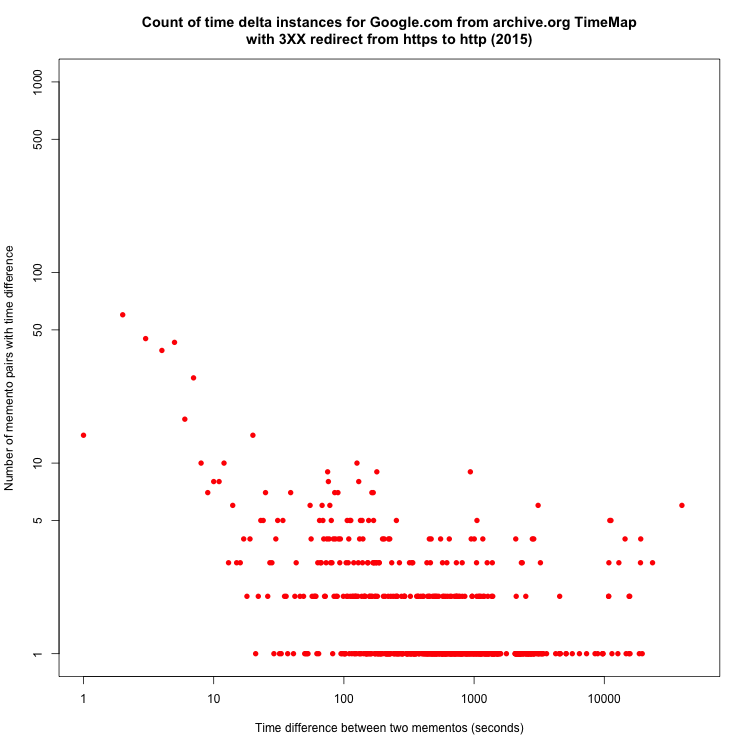}
\end{figure}

\begin{figure}
\centering
\includegraphics[width=1.0\linewidth]{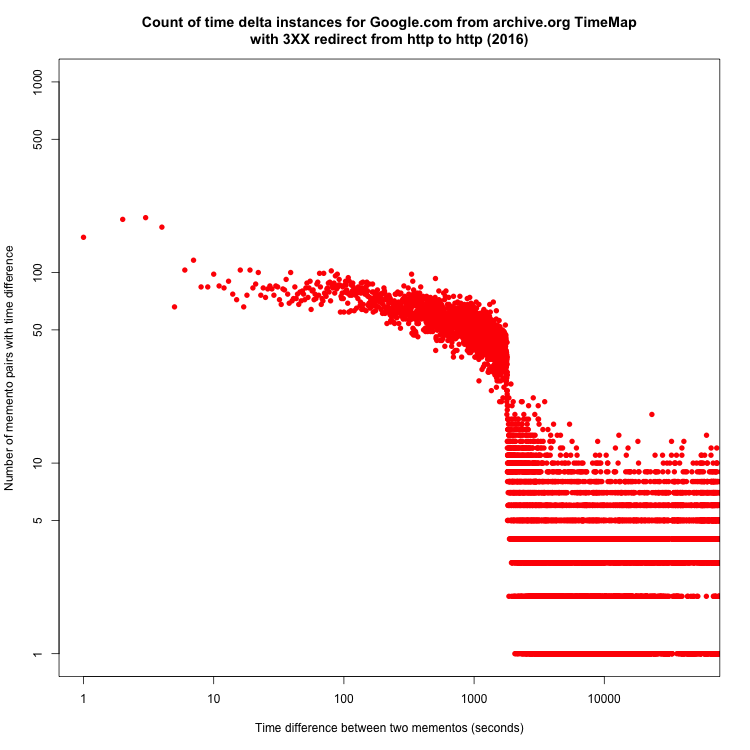}  
\end{figure}

\begin{figure}
\centering
\includegraphics[width=1.0\linewidth]{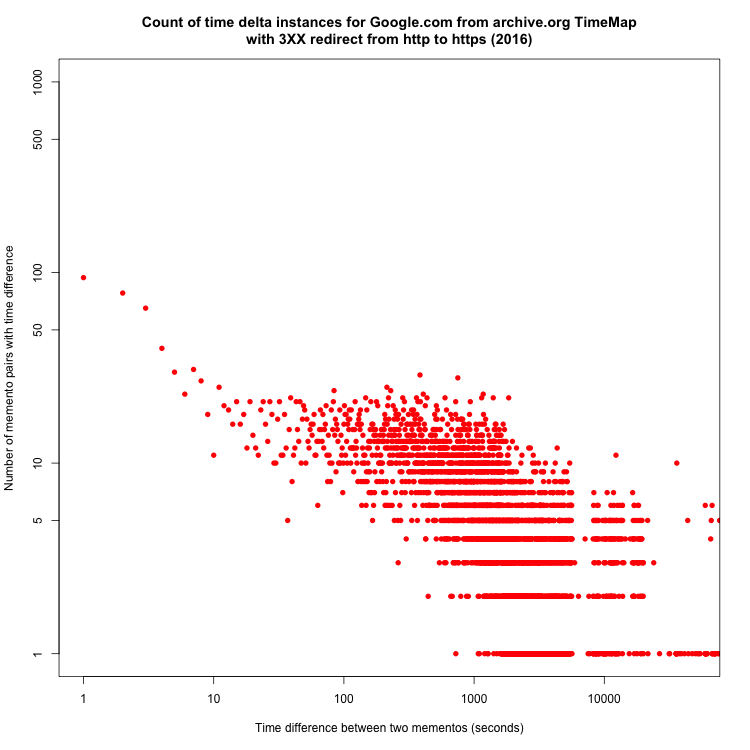}  
\end{figure}

\begin{figure}
\centering
\includegraphics[width=1.0\linewidth]{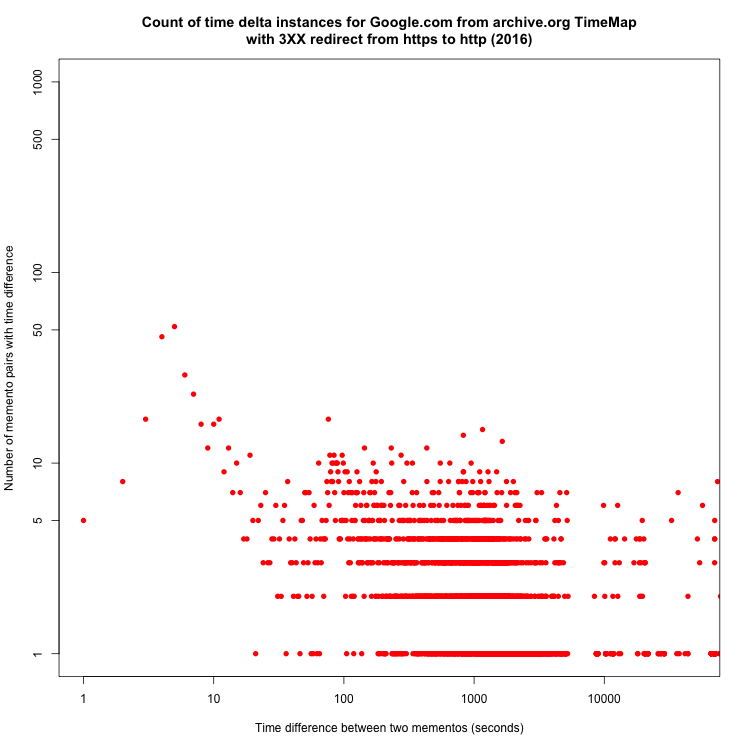}  
\end{figure}

\begin{figure}
\centering
\includegraphics[width=1.0\linewidth]{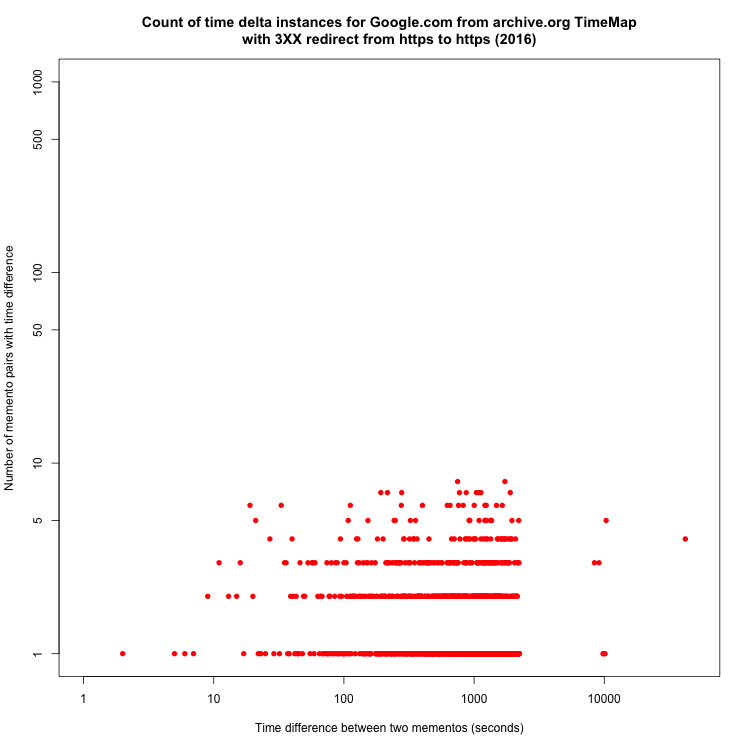} 
\end{figure}

\end{document}